\documentclass[a4paper,12pt]{article}
\usepackage{authblk}
\usepackage{multicol}
\usepackage{amsmath,amssymb}

\usepackage{amsthm} 
\usepackage{hyperref}
\usepackage{amsfonts}
\usepackage{verbatim}
\usepackage{enumitem}
\usepackage{stmaryrd}
\usepackage{graphicx}
\usepackage{svg}
\graphicspath{{./main_img/}{./si_img/}}
\svgpath{{./main_img}{./si_img}}
\usepackage{bbm}
\usepackage{lscape}
\usepackage{multirow}
\usepackage{makecell}
\DeclareUnicodeCharacter{3BA}{$\kappa$}
\usepackage{rotating}
\usepackage{tabularray}
\usepackage{makecell}
\usepackage[a4paper, total={6.5in, 9.5in}]{geometry}
\usepackage{fancyhdr}
\usepackage[style=numeric, abbreviate=true, giveninits=true, terseinits=true,maxnames=10, doi=false, isbn=false, url=false, eprint=true, date=year]{biblatex}
\newtheorem{theorem}{\bf Theorem}[section]

\newtheorem{corollary}{\bf Corollary}[section]
\newtheorem{proposition}{\bf Proposition}[section]
\usepackage{palatino}
\usepackage{hyperref}
\usepackage{amsfonts}
\usepackage{verbatim}
\usepackage{enumitem}
\usepackage{graphicx}
\usepackage{url}

\numberwithin{equation}{section}
\numberwithin{theorem}{section}

\newcommand{\R}{\mathbb{R}}

\newcommand{\bx}{{\bf{x}}}
\newcommand{\by}{{\bf{y}}}
\newcommand{\bz}{{\bf{z}}}
\newcommand{\br}{{\bf{r}}}

\definecolor{heavyred}{cmyk}{0,1,1,0.25}
\hypersetup{
  pdftitle=Title,
  pdfpagemode=UseOutlines,
  citebordercolor=0 0 1,
  colorlinks=true,
  allcolors=heavyred,
  breaklinks=true,
  pdfauthor={Add authors here},
  pdfpagetransition=Dissolve
}

\makeatletter
\newcommand{\spx}[1]{%
  \if\relax\detokenize{#1}\relax
    \expandafter\@gobble
  \else
    \expandafter\@firstofone
  \fi
  {^{#1}}%
}
\makeatother

\newcommand{\genericdel}[4]{%
  \ifcase#3\relax
  \ifx#1.\else#1\fi#4\ifx#2.\else#2\fi\or
  \bigl#1#4\bigr#2\or
  \Bigl#1#4\Bigr#2\or
  \biggl#1#4\biggr#2\or
  \Biggl#1#4\Biggr#2\else
  \left#1#4\right#2\fi
}

\let\abs\envert
\newcommand{\sVert}[1][0]{%
  \ifcase#1\relax
  \rvert\or\bigr|\or\Bigr|\or\biggr|\or\Biggr
  \fi
}

\graphicspath{ {./img/} }
\usepackage{subcaption}
\usepackage{bbm}

\author[1]{K. J. Painter} 
\author[2]{T. Hillen}
\author[3]{J. R. Potts}

\affil[1]{DIST, Politecnico di Torino, Torino, Italy, Viale Pier Andrea Mattioli, 39, 10125.  kevin.painter@polito.it}
\affil[2]{Department of Mathematical and Statistical Sciences, University of Alberta, Canada, thillen@ualberta.ca}
\affil[3]{School of Mathematics and Statistics, University of Sheffield, UK, j.potts@sheffield.ac.uk}

\date{}
\begin{document}

\title{Biological Modelling with Nonlocal Advection Diffusion Equations}
\maketitle

\begin{abstract}
 The employment of nonlocal PDE models to describe biological aggregation and other phenomena has gained considerable traction in recent years. For cell populations, these methods grant a means of accommodating essential elements such as cell adhesion, critical to the development and structure of tissues. For animals, they can be used to describe how the nearby presence of conspecifics and/or heterospecifics influence movement behaviour. In this review, we will focus on classes of biological movement models in which the advective (or directed) component to motion is governed by an integral term that accounts for how the surrounding distribution(s) of the population(s) impact on a member’s movement. We recount the fundamental motivation for these models: the intrinsic capacity of cell populations to self-organise and spatially sort within tissues; the wide-ranging tendency of animals towards spatial structuring, from the formations of herds and swarms to territorial segregation. We examine the derivation of these models from an individual level, illustrating in the process methods that allow models to be connected to data. We explore a growing analytical literature, including methods of stability and bifurcation analysis, and existence results. We conclude with a short section that lays out some future challenges and connections to the modelling of sociological phenomena including opinion dynamics.
\end{abstract}


\section{Introduction}

A flamboyance of flamingos, a shiver of sharks, a confusion of wildebeest; hundreds of collective nouns have been assigned to define the groups formed by different species. The need for these collective nouns reflects the frequency with which animal groups form across the natural world, from the gathering of a small number of individuals to billions-strong swarms of locusts \cite{roussi2020} or a herring shoal that stretches across kilometres \cite{makris2009}. An ability to aggregate is a phenomenon that extends down to the microscopic level, where various bacteria \cite{budrene1991,budrene1995} and microorganisms\cite{bonner2009} have been observed to organise into aggregates under certain conditions. In the context of our own cells, their capacity to bind and organise is key for the development of many tissues and organs, or their repair following injury. 

An essential element in the formation of many groups is the triggering of a movement-based response in an individual, according to signals and  behaviours of other members. Directly, a cell may touch another cell and pass information through specialised molecules at the cell surface, or a bird may alter its flight path according to the trajectory of a neighbour. Indirectly, cells may alter motility according to a molecular signal deposited by another cell and animals may respond to territorial scent markings of conspecifics. The cumulative effect of these individual-level behaviours can result in self-organisation at the population scale, for example the rounding up of an initially dispersed population into an aggregate or the adoption of some swarm configuration.

Scientific interest in self-organising phenomena has a long history, and the field forms a pillar of mathematical biology \cite{murray2003}. Naturally, much of the modelling within this field is indebted to the remarkable work\cite{turing1952} of Alan Turing through his reaction-diffusion model, proposed to explain how morphogenesis could occur. Turing's model involved only molecular components, and showed how an interplay between reaction and diffusion could break the symmetry of a spatially uniform distribution by amplifying natural stochastic fluctuations into an ordered and patterned state. This not only offered a plausible chemical blueprint for how a tissue could become patterned, but also a mathematical blueprint for determining whether self-organisation can occur in some system. Inspired by the aggregation mounds formed from starving {\em Dictyostelium discoideum} cells -- the initiating step during a multicellular transformation that serves as a paradigm of self-organisation at the microscopic scale \cite{bonner2009} -- the celebrated chemotaxis model of Keller and Segel \cite{keller1970} followed Turing's template to illustrate how a system that includes an actively migrating population could also undergo self-organisation. It shows that the positive feedback loop of chemotaxis to a self-secreted attractant could lead to mound formation. 

Continuous biological movement models are often formulated as an advection-diffusion equation \cite{murray2003}, i.e.
\begin{equation}\label{diffusionadvection}
	\partial_t u({\bf{x}},t) = \nabla \cdot \left[ D \nabla u({\bf{x}},t) - {\bf{a}} u({\bf{x}},t)\right]\,,
\end{equation}
where $u({\bf{x}},t)$ represents the density of some population at position ${\bf{x}}\in \Omega \subset \R^n$ and time $t\in [0,\infty)$. $D$ measures the diffusive (undirected) component to movement, while ${\bf{a}}$ is an $n$-dimensional vector that measures the advective (directed) component to movement. Generally, diffusion may be an  $n\times n$ diffusion tensor matrix, e.g. describing some anisotropic spread due to the environment \cite{hillen2013}, however here we will generally take an isotropic diffusion represented by a scalar coefficient $d$, so that $D=dI_n$ where $I_n$ is the $n \times n$ identity matrix. The region $\Omega$ defines the space in which the population moves: this could range from a line if movement is effectively constrained to a one-dimensional geometry ($n=1$, e.g. cell movement along nano-engineered channels), a two-dimensional surface ($n=2$, e.g. animal movement across a landscape) to a three dimensional volume ($n=3$). If $\Omega$ is a bounded domain, then the above model (\ref{diffusionadvection}) will be equipped with appropriate boundary conditions. 

For the chemotaxis model of Keller and Segel \cite{keller1970} interactions between individuals are indirect: the individual senses (and moves in response to) another individual through following the local gradient of an attractant secreted by the population.
 As such, the advective velocity is taken to be proportional to the chemoattractant gradient, i.e. $ {\bf{a}} \propto \nabla v$, where $v$ is the attractant.
 
 In other instances of group formation, however, interactions are direct: molecular binding between receptors on adjacent cell surfaces can lead to cells pulling themselves together (adhesion or attraction) or moving away from each other (repulsion); animals may also be drawn to each other or move away following a visual sighting of conspecifics. In all such instances, the interaction range becomes a crucial point for consideration: in the case of cells, this could be the range over which a cell can contact a neighbouring cell through touch, or, for animals, the range over which the perception of conspecifics influences its movement behaviour. 

Given the existence of an interaction range, an individual has the potential to sense multiple neighbours simultaneously. It is natural, therefore, to suppose that the movement will be based on some integrated response, i.e. according to the distribution of a population (or populations) across its interaction range. Such considerations have led to the increasing adoption of nonlocal PDE formulations \cite{chen2020}. The focus of attention in the present review will be on models in which the nonlocality appears within the advective term, which is calculated according to an integral that measures the influence of the surrounding population on  movement. Specifically, we consider the following pair of non-local models,
\begin{subequations} \label{aggregation}
\begin{align}
\partial_t u & = d\Delta u -\mu \nabla \cdot \left[ u {\bf k}_R \ast  f  \right], &
{\bf k}_R  \ast f \,({\bf{x}},t)  = \int_{\Omega} {\bf {k}}_R ({\bf{x}},{\bf{y}}) f(u({\bf{y}},t)) d{\bf y}\,,
\label{aggregation1}\\
\partial_t u & = d\Delta u -\nu \nabla \cdot \left[u \nabla ({w}_R \ast g) \right], &
{w}_R \ast g \,({\bf{x}},t) = \int_{\Omega} w_R ({\bf{x}},{\bf{y}}) g(u({\bf{y}},t)) d{\bf y} \,.\label{aggregation2}
\end{align}
\end{subequations}
Motivation for these two model forms can be found through a purely phenomenological argument or by applying a more physical-based reasoning.

\begin{figure}[t!]
    \centering
    \includegraphics[width=\textwidth]{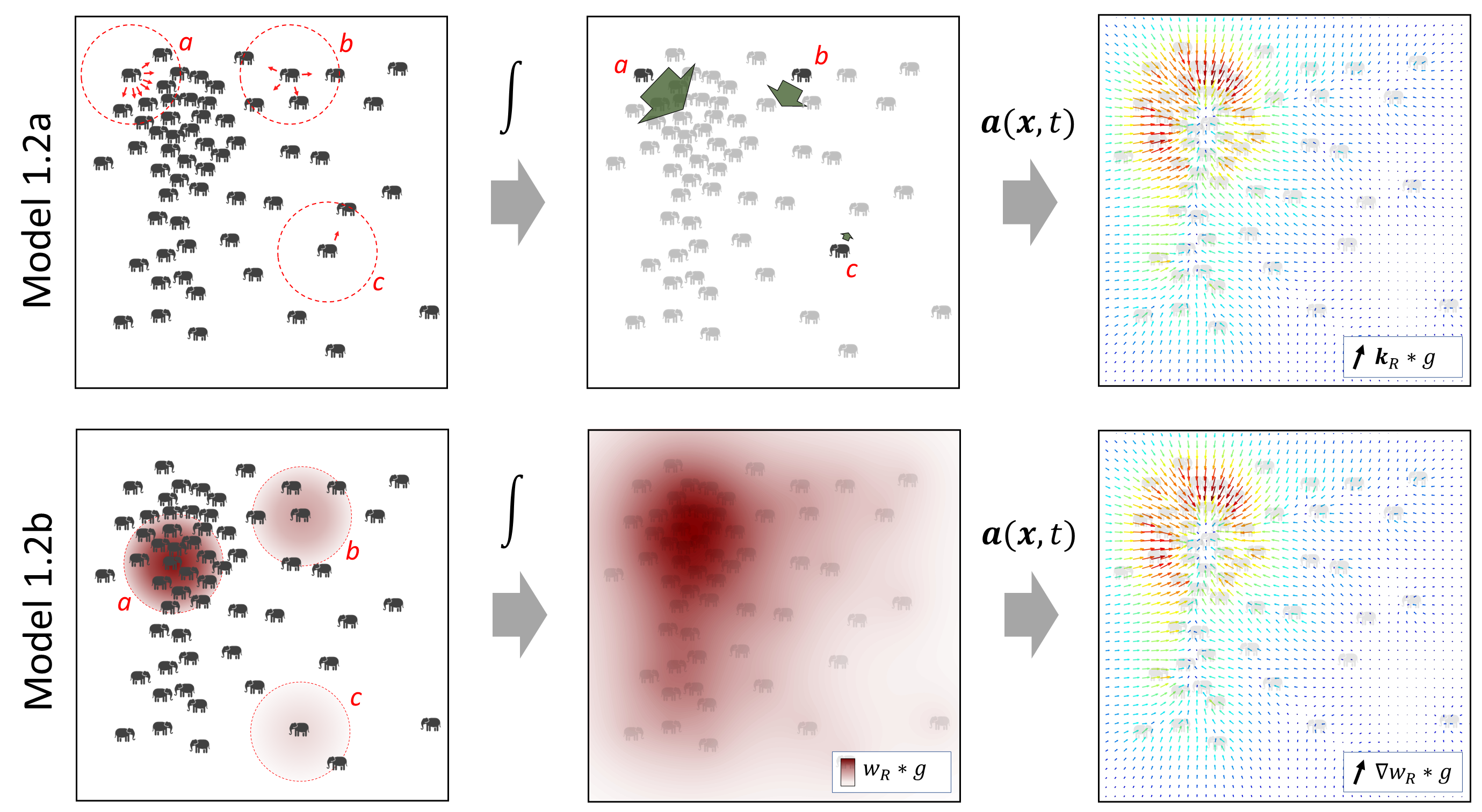}
    \caption{Illustration of the models (\ref{aggregation}) as formulated to describe grouping or herding, i.e. a tendency to move towards and aggregate at areas of higher population density. (Top row) For \eqref{aggregation1} each individual within the interaction region (dotted circles) generates a local `force' of attraction (top left); the number and direction will be different according to each individual's position (points a,b,c). Integrating over the interaction region leads to a net movement, the strength and direction varying with position (top middle). Overall, this generates an advective field that directs population level movement (top right). (Bottom row) For \eqref{aggregation2} an individual measures the (nonlocal) population density, e.g. by assessing the number of neighbours within the interaction region; at distinct positions (a,b,c), different numbers of neighbours will be detected (bottom left). Across space, this creates a population distribution map (bottom middle). The advective field for the population is according to the gradient of this distribution map (bottom right), e.g. in the direction of increasing gradient to describe a herding phenomenon. The advection fields generated through these two formulations have a similar form.}
    \label{schematic}
\end{figure}

Consider first the formulation \eqref{aggregation1} and its phenomenological motivation (see top row of Figure \ref{schematic}). Here, the nonlocal advection term is founded on the principal that the population at position ${\bf y}$ influences the movements of those at ${\bf x}$.  The induced direction of movement and its magnitude depends on the product of a (vector-valued) function ${\bf k}_R ({\bf{x}},{\bf{y}})$ and a (scalar-valued) function $f(u({\bf{y}},t))$. Specifically, ${\bf k}_R ({\bf{x}},{\bf{y}})$ specifies a dependence on the distance of ${\bf y}$ to ${\bf x}$ and it identifies the direction of interaction. The function  $f(u({\bf{y}},t))$ defines the dependence on the population size at ${\bf{y}}$. The integral kernel ${\bf k}_R$ is parametrised according to a sampling radius $R$, representing the interaction range. Net movement results from integrating over all possible positions, and this directly informs the advective velocity at ${\bf x}$. The parameters $d \in \R^+$ and $\mu \in \R$ describe diffusion and advection coefficients, respectively.

The phenomenological motivation for \eqref{aggregation2} follows a similar reasoning (see bottom row of Figure \ref{schematic}), although the function $w_R ({\bf{x}},{\bf{y}})$ is now scalar-valued, as is the integrated quantity ${w}_R \ast g$. This formulation can be interpreted analogously to the taxis-like model, with the population moving according to the gradient of a nonlocal measure of the population; for example, this could be a nonlocally-averaged density distribution. Again, the parameters $d \in \R^+$ and $\nu \in \R$ represent diffusion and advection coefficients, respectively.

A physical reasoning for \eqref{aggregation1} and \eqref{aggregation2} follows the consideration of forces and energies; this interpretation takes on particular resonance in the context of cell migration, where translocation of a cell's body stems from forces exerted as it attaches to other cells and the substrate. Model (\ref{aggregation1}) can be derived through a balance between adhesive and repulsive forces that act at the cell surface (e.g. see \cite{buttenschon2018}): interactions between cells centred at ${\bf x}$ and ${\bf y}$ generate local forces, with the net force according to the integral ${\bf k}_R \ast f$. This quantity hence describes a force density, with units of $N/m$, and the coefficient $\mu $ has units of $(s N)^{-1}$. The corresponding term for (\ref{aggregation2}) is  $\nabla (w_R \ast g) $, and $w_R\ast g$ will carry the units of an energy density ($J/m$). Viewed in this light, the advection according to $\nabla (w_R \ast u) $ defines a movement according to an energy gradient. The summary review of  \cite{Carrillo-chapter} describes the derivations of models (\ref{aggregation2}) according to energy principle. If the underlying principal is a process of energy minimisation (i.e. down the energy gradient) then parameter $\nu <0$ and, conventionally, the form (\ref{aggregation2}) is written with the sign of the advection term reversed, i.e. 
\begin{equation} \label{aggregation2alt}
\partial_t u = d\Delta u +\gamma \nabla \cdot \left[u \nabla ({w}_R \ast g) \right]\,, \quad\quad 
{w}_R \ast g \,({\bf{x}},t) = \int_{\Omega} w_R ({\bf{x}},{\bf{y}}) g(u({\bf{y}},t)) d{\bf y} 
\end{equation}
where $\gamma > 0$ indicates energy minimisation. Note that we will adopt this convention particular in Section \ref{s:existence}-\ref{s:bifurcation}, where energy-based analytical methods are utilised. Since energy differences lead to forces, a natural connection between these model forms is laid bare. A note of caution, though, must be applied when applying physical reasoning to biological particles such as animals or cells: attraction between conspecifics or avoidance of predators are measurable behaviours, but they cannot be directly related to a physical force or energy; similarly, a cell is a highly complex structure and its behaviour is not necessarily determined by the need to minimise energy. 

Models of form \eqref{aggregation} have been used since the 1970s to describe ecological systems (see \cite{kawasaki1978,mimura1982,grunbaum1994,mogilner1999,sekimura1999,lee2001,okubo2001,topaz2006}), since the 1990s to describe cellular systems (see \cite{sekimura1999,armstrong2006,gerisch2008,murakawa2015,carrillo2019}), and more recently to describe opinion dynamics (see \cite{garnier2017,goddard2022}). A particular point of mathematical interest lies in their capacity for self-organisation, in which modelling a process of self-to-self attraction between members can allow a dispersed population to organise itself into one or more aggregated groups. For this reason, they are commonly referred to as aggregation equations. However, it is important to note that the formulations \eqref{aggregation} are less restrictive and can be used to model other forms of interaction, such as repulsive interactions that could lead to an enhanced dispersal. 

Moreover, the form of these models can be extended to describe heterogeneous populations where the interactions between different populations can be distinct (e.g. see \cite{armstrong2006,murakawa2015,painter2015,pottslewis2019}) or incorporated within more complicated models and applied to explain specific phenomena, such as cancer invasion for cellular systems (e.g. see \cite{gerisch2008,painter2010,domschke2014}) or dynamics of locust swarms in ecological systems (see \cite{topaz2012,georgiou2021}). A multi-species generalisation of each of the models (\ref{aggregation1}-\ref{aggregation2}) can easily be formed by extending to ${\bf u}({\bf x},t)=(u_1({\bf x},t), \hdots, u_p({\bf x},t))$, where $u_i$ denotes the density distribution of the $i^{th}$ out of $p$ populations, and considering the systems
\begin{subequations} \label{multiaggregation}
\begin{align}
	\partial_t u_i =   d_i\Delta u_i & - \sum_{j=1}^{p}\mu_{ij} \nabla \cdot \left[ u_i {\bf k}_{ij} \ast  f_{ij}  \right] \label{multiaggregation1} \\
	& {\bf k}_{ij} \ast f_{ij}  = \int_{\Omega} {\bf {k}}_{ij} ({\bf{x}},{\bf{y}}) f_{ij}({\bf u}({\bf{y}},t)) d{\bf y}  \quad i = 1 \hdots p\,, \nonumber\\
	\partial_t u_i = d_i\Delta u_i & -\sum_{j=1}^{p}\nu_{ij} \nabla \cdot \left[u_i \nabla ({w}_{ij} \ast g_{ij} )\right] \label{multiaggregation2} \\
	& {w}_{ij} \ast g_{ij} = \int_{\Omega} w_{ij} ({\bf{x}},{\bf{y}}) g_{ij}({\bf u}({\bf{y}},t)) d{\bf y} \quad i = 1 \hdots p \,. \nonumber
 \end{align}
\end{subequations}
In model (\ref{multiaggregation1}) directed movement is now the combined result of $N$ movement-inducing interactions, where ${\bf k}_{ij} \ast  f_{ij}$ is the nonlocal advection coefficient that defines the movement induced on members of population $i$ due to interactions with population $j$: ${\bf {k}}_{ij}({\bf x},{\bf y})$ and $f_{ij}({\bf u}({\bf{y}},t))$ are analogous to the functions described above, and parameters $R_{ij}$, $d_i$, and $\mu_{ij}$ define the interaction range, diffusion coefficient and advection coefficients, respectively. Note that the $\mu_{ij}$'s may be positive or negative, to model inter-species \cite{pottslewis2019} or inter-cellular \cite{painter2015} attraction or repulsion, respectively. Analogous reasoning can be applied to the form (\ref{multiaggregation2}).

In this article we review the increased employment of nonlocal systems of the above form within biological modelling\footnote{Given the scope of this article, we cannot cover all topics in detail and many relevant studies are ommitted.}. In Sections 2 and 3 we outline our motivating biological systems, namely cellular adhesion and other cell-based interactions (Section 2) and ecological interactions between animals (Section 3). We describe the key biology and previous modelling that has motivated models of the form (\ref{aggregation1}) and 
(\ref{aggregation2}) or their multiple species extensions. In Section 4 we explore the derivations of these models from a microscopic perspective, in particular focussing on cellular adhesion. In Section 5 we consider some of the analysis used to understand these models, including linear stability analysis, bifurcation analysis and global existence. We conclude with some key challenges and future perspectives for the field.

\section{Nonlocal models for cellular systems}

\subsection{Adhesion and other cell interactions}

\begin{figure}[t!]
	\includegraphics[width=\textwidth]{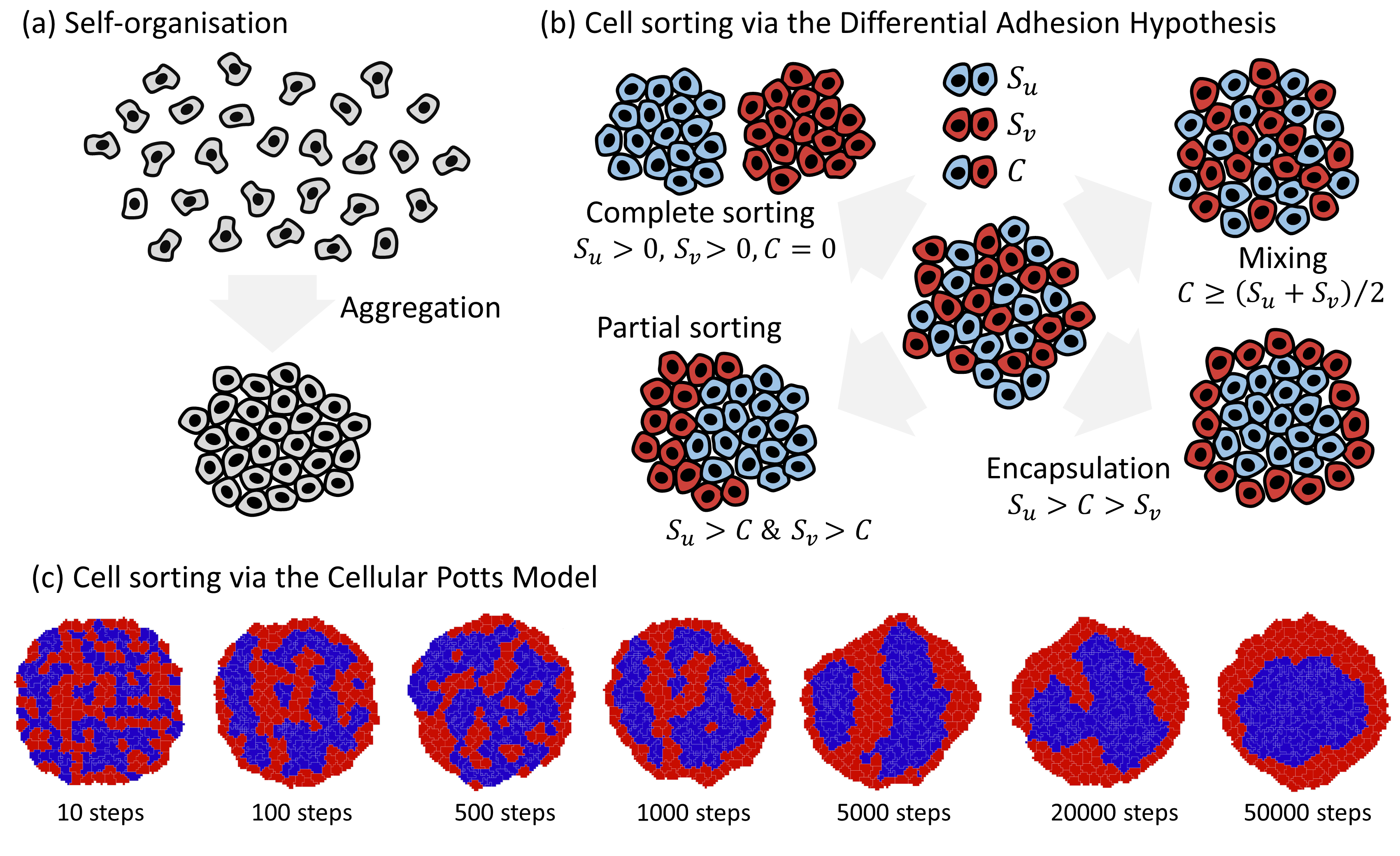}
	\caption{(a) Cell-cell adhesion naturally leads to accretion, with cells attaching on contact and forming a cluster or aggregation. (b) Sorting dynamics in adhesive populations, as predicted by the DAH. In a mixture of two distinct cell populations, three principal parameters can be identified: two self-adhesion strengths ($S_u$, $S_v$) and one cross-adhesion strength ($C$). The DAH predicts that different arrangements will arise according to the relationship between these parameters: for example, in a mixture of cells in which $S_u > C > S_v$, the $u$ population (red) becomes encapsulated by the $v$ (blue) population. (c) CPM simulation (implemented via Compucell3D) showing encapsulation for a parameter setting in which adhesive interactions satisfy the aforementioned relationship.} \label{aggregationsorting}
\end{figure}
 
Cell adhesion is the fundamental mechanism by which a cell attaches to and interacts with its surroundings\cite{alberts2015}. Adhesions form through specialised cell surface receptors; their binding across adjacent membranes not only attaches cells together, but also triggers a range of processes from proliferation to migration. Of the various families of adhesion molecules, cadherins play a particularly prominent role within cell-cell adhesion processes (e.g. see \cite{takeichi2022}): E-cadherins, for example, form tight adhesive junctions between epithelial cell types; N-cadherins are more commonly associated with transient adhesive interactions between motile mesenchymal cells. 

Adhesion is critical for the organisation and maintenance of tissue structure. Naturally, cell-cell adhesion can lead to an accretion process, whereby contact between cells leads to attachment and the formation of a clustered population, Figure \ref{aggregationsorting}(a). Moreover, classic experiments indicate a role for adhesion in regulating the spatial organisation of different populations within a tissue \cite{townes1955}. In the differential adhesion hypothesis (DAH)\cite{steinberg2007,tsai2022} cell sorting is suggested to result from distinct cell surface tensions, deriving in turn from the strength of adhesive interactions. The precise relationship leads to different configurations, see Figure \ref{aggregationsorting}(b), and experiments\cite{foty2005} for cell lines that express different levels of cadherins are consistent with this theory. More recently, measurements of the forces within adhesive aggregates  \cite{amack2012,tsai2022} have resulted in revision of the DAH to the differential interfacial tension hypothesis (DITH \cite{brodland2002}): cell cortical contraction machinery and cell-cell adhesion combine to regulate interfacial tension, and sorting results from rearrangements that lead to a tissue-level minimisation of interfacial tension. Nevertheless, adhesion remains the driving force within the sorting and arrangement of tissues.

Cell-to-cell contacts, though, can also trigger repulsion. For example, contact inhibition of locomotion (CIL)\cite{abercrombie1954} forms a contact-mediated response which not only leads to cessation of cell motion, but also repolarisation and reversal of the direction of motion \cite{carmona2008}. Cell-to-cell contacts can also lead to asymmetric responses, where the two cells display contrasting responses. One such example arises in the pigmentation of zebrafish, where interacting xanthophores and melanophores engage in a chase and run\cite{inaba2012,yamanaka2014} interaction, contact between them resulting in the melanophore moving away from the pursuing xanthophore. Other instances of contact-mediated responses that can range from attraction to repulsion include those triggered through Eph/Ephrin interactions \cite{bush2022} or the chase and run dynamics observed in cultures of neural crest and placode cells \cite{theveneau2013}. A complex set of migration responses that follow direct contacts have been observed among cells of the immune system, impacting on a range of processes that include inflammation and tumour progression \cite{miskolci2021}.

Biological cells are small with an average diameters the order of around ten microns and contact-based interactions occur at a similarly local level. However, contacts can also be formed at considerably greater distances than the mean cell diameter. First, the cell bodies can be highly deformable, where frequent protrusions of the membrane -- pseudopodia \cite{chodniewicz2004} -- locally extend parts of the membrane far beyond the average diameter. Second, a diversity of more specialised membrane protrusions have been identified \cite{yamashita2018,korenkova2020,roehlecke2020} -- variously termed cytonemes, tunnelling nanotubes, microtubes -- that in some cases extend the order of 100s of microns. Thus, a contact can be achieved between cells separated by multiple cell diameters, and a non-local description is warranted.

\subsection{Models for adhesion and tissue dynamics} 

\subsubsection{Individual level models for adhesion and sorting} \label{s:discreteadhesion}

Agent-based modelling (ABM) forms a natural approach for adhesive cell populations\cite{steinberg1975,sulsky1984}. The first broadly successful {\em in silico} replications of cell sorting can be attributed to Graner and Glazier \cite{graner1992,glazier1993}, where a Potts model\footnote{A model of statistical mechanics, originally used to understand spin configurations in ferromagnets.} was extended to model adhesion. Subsequently dubbed the Cellular Potts Model (CPM), each biological cell occupies multiple grid cells spread across a lattice, therefore giving each cell a shape, volume, and boundary. Evolution of the shape is probabilistically determined via a hypothesised energy functional; the aim is to minimise an energy determined by adhesive contacts along shared surfaces. Selecting relationships in line with the DAH leads to the predicted cell sorting pattern \cite{graner1992,glazier1993}; see Figure \ref{aggregationsorting}(c) for a CPM simulation in which adhesion relationships conspire to sort two populations into an encapsulated configuration.

Other ABMs have also shown to be capable of describing adhesion and sorting dynamics\cite{van2015}, sitting at various levels of detail: cells modelled as deformable ellipsoids  \cite{palsson2000,palsson2008} with centres and semi-axes evolving according to the forces generated by adhesive interactions with other cells and the substrate; on-lattice methods, (e.g. cellular automata type, see \cite{deutsch2005}); off-lattice centre-based models, where equations of motion describe the position and velocity of a cell's centre and the cell forms a hard or soft sphere that interacts with nearby cells (e.g.  \cite{hoehme2010,macklin2012,carrillo2018}); vertex-based models \cite{fletcher2014} which feature cell boundaries described by a polyhedron with dynamic vertices. Many of these ABMs form the basis of computational platforms for simulating cellular and tissue dynamics -- CellSys\footnote{\texttt{https://www.hoehme.com/software/tisim}} \cite{hoehme2010}, CompuCell3D 
\footnote{\texttt{https://compucell3d.org/}} \cite{swat2012}, Chaste \footnote{\texttt{https://www.cs.ox.ac.uk/chaste/}} \cite{mirams2013}, Physicell\footnote{\texttt{http://physicell.org/}}, \cite{ghaffarizadeh2018} -- and their capacity to predict adhesion and sorting phenomena is regarded as a point of calibration between these diverse methodologies \cite{osborne2017}.

\subsection{Continuous models for adhesion and sorting}

\subsubsection{Local formulations}

The representation of a cell population via a continuous density distribution eliminates the issue of scale inherent to agent-based models, where simulating very large cell numbers remains a computational challenge. Moreover, a well posed differential equation system gives access to a wealth of analytical methods (stability and bifurcation analysis, asymptotic approaches, travelling wave analyses) that can yield deeper understanding into the dynamics. 

One simple approach to include adhesion has been based on a classic advection-diffusion equation of the form (\ref{diffusionadvection}), where the diffusion and/or advection coefficients depend on the  local population density, i.e. the pointwise density. Such models have been proposed on phenomenological grounds (e.g. see \cite{hofer1995}), or following a derivation from an underlying random walk description of movement (e.g. see \cite{anguige2009,johnston2012}) -- see Section \ref{s:randomwalks}. These models capture certain features of adhesive populations -- for example, restricted motility in regions of high adhesiveness -- and are both analytically straightforward and simple to incorporate into models. Nevertheless, they have not been shown to allow more complicated sorting behaviour. Moreover, as discussed in greater detail below, the derived diffusion coefficients can sometimes become negative and result in a loss of regularity (for example, \cite{anguige2009,johnston2012}). The effects of cell-cell adhesion have also been incorporated in a phenomenological manner into various models for tumour growth (for example \cite{byrne1996,byrne1997,cristini2003,wise2008,cristini2009}), via the incorporation of a surface tension force at the tumour-tissue surface.

\subsubsection{Nonlocal formulations}

Successful ABM approaches for cell sorting are inherently nonlocal: a cell spread across multiple lattice sites in a CPM, or centre-based approaches where the attractive and repulsive interactions form over an interaction range. This nonlocality can be incorporated into a continuum description using a nonlocal (or integral) PDE formulation. In the context of cell adhesion, the first\footnote{As far as we are aware} models to adopt this approach were formulated  to describe the aggregation of a single homogeneous population in \cite{sekimura1999} and for multiple cell populations in\cite{armstrong2006} to explore sorting via differential adhesion; closely related nonlocal models, though, have a biomodelling history that dates back at least as far as the 1970s (for example, see \cite{kawasaki1978,mimura1982,nagai1983,grunbaum1994}). 

The simplest motivation for these models is founded on phenomenological reasoning. Suppose $u({\bf x},t)$ denotes the cellular density at position ${\bf x}$ in space and $t$ in time. Ignoring (for simplicity) cellular growth or death and employing standard mass conservation arguments (e.g. see \cite{murray2003}) leads to the balance equation 
\[
\partial_t u({\bf x},t) = -\nabla \cdot {\bf J}({\bf x},t)\,,
\]
where ${\bf J}({\bf x},t)$ denotes the population flux arising from movement. The flux can be decomposed into different terms -- for example, a diffusive element to describe undirected movement and an advective component for directed movement -- and we arrive at \eqref{diffusionadvection}. Regarding the advective component, suppose that a cell at ${\bf x}$ interacts with another cell at ${\bf y}$, and that this interaction generates movement; this could be the result of forming adhesive bonds that draw the two cells together. The net movement response follows from summing over all possible interactions and we then postulate an interactive flux proportional to this sum, i.e. 
\[
{\bf J}_{\mbox{{\it interaction}}} \propto u({\bf{x}},t) \int {\bf {k}}_R ({\bf{x}},{\bf{y}}) f(u({\bf{y}},t))\, d{\bf{y}}\,.
\]
where ${\bf {k}}_R$ and $f(u({\bf{y}},t))$ are as described following (\ref{aggregation}). Adding to the above a standard (Fickian) diffusive flux, ${\bf J}_{\mbox{{\it diffusion}}} = -d \nabla u$, leads to \eqref{aggregation1}. 

A basic model to describe a homogeneous adhesive population sets ${\bf {r}} = \bf{y}-\bf{x}$,
\begin{equation}\label{kfforms}
{\bf {k}}_R ({\bf{x}},{\bf{x}}+{\bf{r}}) = \chi_{\left|{\bf{r}}\right|<R} \vec{\bf{e}}_r 
\quad \mbox{and} \quad f(u({\bf{x}}+{\bf{r}},t)) \propto u({\bf{x}}+{\bf{r}})\,,
\end{equation}
where $\vec{\bf{e}}_r $ denotes the unit vector in direction of ${\bf{r}}$, and $\chi({\bf {r}})$ is the indicator function. This stipulates (i) that only those cells within an interaction range $R$ impact on movement, i.e. those within contact range for adhesive binding; and (ii) that the strength of interaction increases linearly with the density of cells at ${\bf{x}}+{\bf{r}}$, since a higher cell density implies a greater likelihood of forming adhesive bonds. Consequently, we obtain
\begin{equation}\label{simpleadhesion}
\partial_t u = d\Delta u -\mu(R) \nabla \cdot \left(u \int_{B^n_R} u({\bf{x}}+{\bf{r}},t) \vec{\bf{e}}_r \, d{\bf{r}} \right)\,,
\end{equation}
where $B^n_R $ is the $n$-dimensional ball of radius $R$. The coefficient $\mu > 0$ is a measure of the adhesive strength; switching to $\mu <0$ turns the interaction into a repelling one, e.g. see \cite{painter2015} in the context of CIL. We note that often the function ${\bf {k}}_R$ is normalised, e.g. according to the volume of the interaction space and we therefore place a dependency on $R$ in the parameter $\mu$ for generality. Other natural choices would be to assume that the strength of interaction decreases with increasing separation, due to reduced likelihood of forming a contact: for example, the magnitude of $\bf {k}$ decreasing exponentially with the distance $\left|{\bf{r}}\right|$. Nonlinear choices for $f$ are also logical, e.g. forms to reflect an upper bound in the adhesive pull that can be generated, see below.

\subsubsection{Capacity for self-organisation and sorting} \label{s:aggregation}

\begin{figure}[t!]
    \centering
    \includegraphics[width=\textwidth]{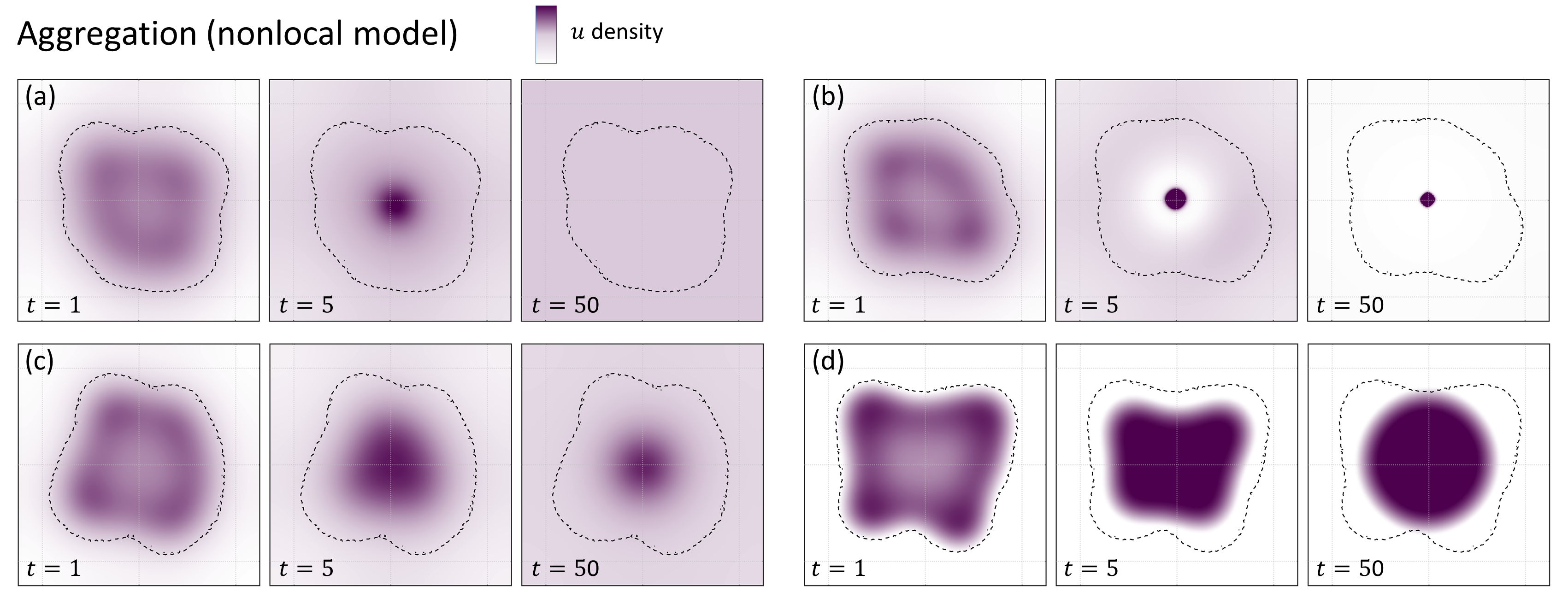}
    \caption{Self-organisation in a nonlocal model for adhesion, homogeneous population. The initial distribution sets a `loose aggregate', the spatial extent of which is indicated by the dashed line in each frame. (a) Dispersal scenario for (\ref{simpleadhesion}), with $d=R=1$ and $\mu = 3.5/\pi$; (b) Aggregation for (\ref{simpleadhesion}), with $d=R=1$ and $\mu = 4/\pi$. (c) Aggregation for (\ref{saturatingadhesion}), for  $d=R=K=1$ and $\mu = 13.5/\pi$; (d) Aggregation for (\ref{saturatingadhesionpressure}), for $d=R=K=1$ and $\mu = 13.5/\pi$. The overall domain $\Omega$ is of size 10$\times$10. We refer to  \protect\cite{gerisch2010,gerisch2010b} for details of the numerical implementation.}
    \label{adhesionsims1}
\end{figure}

A key strength in the model (\ref{simpleadhesion}) lies in its capacity for self-organisation (see Section \ref{s:linearstability} for more details): for $\mu < \mu_{crit}$, a dispersed population remains dispersed, see Figure \ref{adhesionsims1}(a) while for $\mu > \mu_{crit}$ it becomes concentrated into a tight aggregate, see Figure \ref{adhesionsims1}(b). Under the basic model (\ref{simpleadhesion}), the aggregates evolve into a highly concentrated aggregate\footnote{For a discussion of global existence, see Section \ref{s:existence}}, even for $\mu \gtrsim \mu_{crit}$. This can be attributed to the lack of any mechanism that reins in the amount of adhesive pull that can be generated. 

Adding further detail to the model assumptions can help prevent over-accumulation within the aggregates. For example, setting $f(u)$ to be a saturating function (which can be motivated naturally through adhesive receptor occupancy, see Section \ref{sec:deriv}), then 
\begin{equation}\label{saturatingadhesion}
\partial_t u = d\Delta u -\mu(R) \nabla \cdot \left(u \int_{B^n_R} \frac{u({\bf{x}}+{\bf{r}},t)}{\kappa+u({\bf{x}}+{\bf{r}},t)} \vec{\bf{e}}_r \, d{\bf{r}} \right)\,.
\end{equation}
This leads to aggregations that are capped at lower densities, see Figure \ref{adhesionsims1}(c). Other possible modifications include the addition of `volume-filling' (e.g. see \cite{painter2015,carrillo2019}), or adapting diffusion to a density-dependent and degenerate form  (e.g. see \cite{morale2005,burger2007,murakawa2015,burger2018,carrillo2019}). The addition of the latter to (\ref{saturatingadhesion}) leads to
\begin{equation}\label{saturatingadhesionpressure}
\partial_t u = d\nabla \cdot \left[ u \nabla u -\mu(R) \left(u \int_{B^n_R} \frac{u({\bf{x}}+{\bf{r}},t)}{\kappa+u({\bf{x}}+{\bf{r}},t)} \vec{\bf{e}}_r \, d{\bf{r}} \right) \right]\,.
\end{equation}
This adaptation limits a diffusive spread at the cluster boundary, the aggregate taking on a compact form with a sharp interface, Figure \ref{adhesionsims1}(d).

\begin{figure}[t!]
    \centering
    \includegraphics[width=\textwidth]{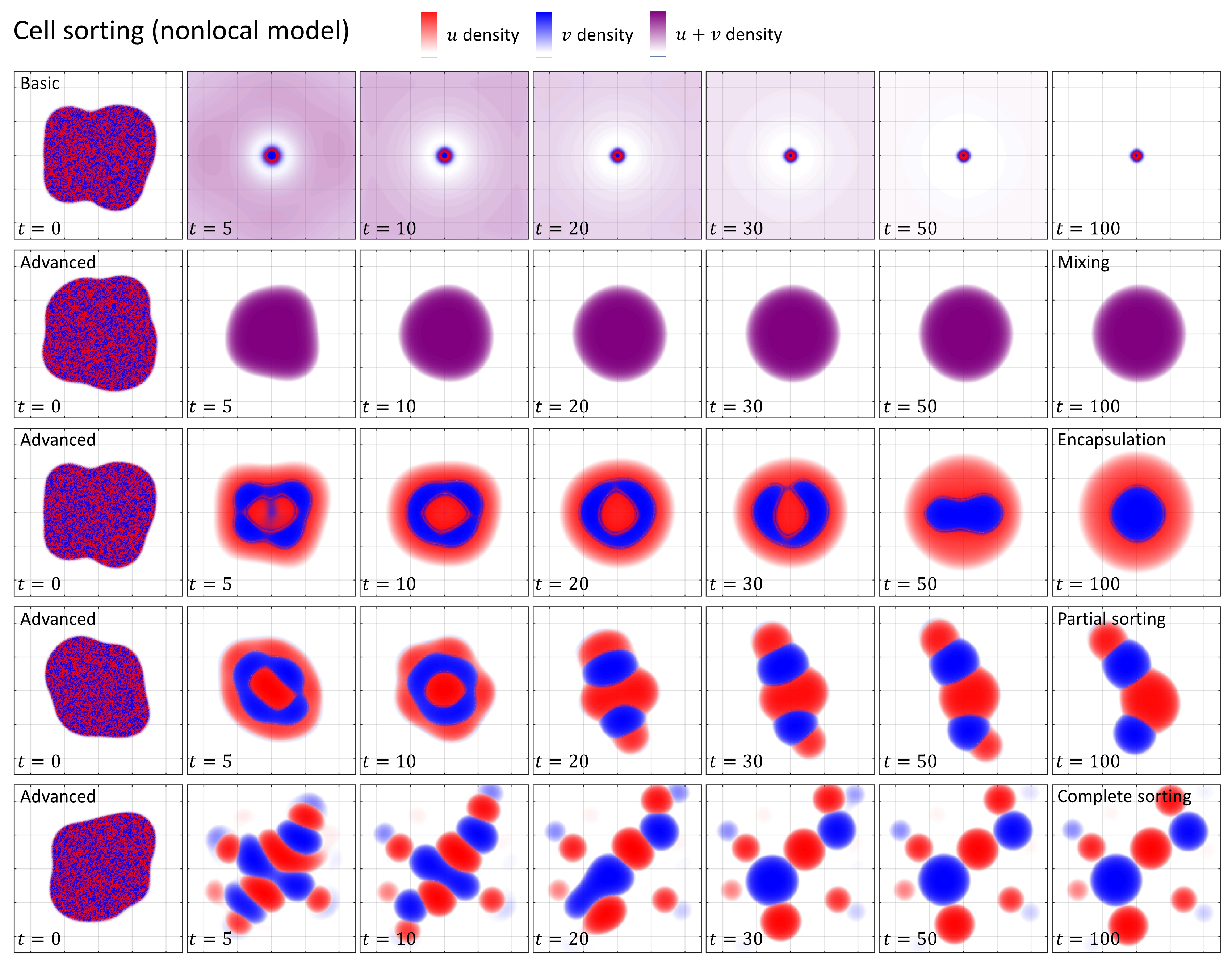}
    \caption{Cell sorting in a nonlocal heterogeneous two population model for adhesion. Initially, the two populations are mixed within a loose aggregate, left column. First row shows a simulation of the basic model (\ref{simpledah}) under $S_u=4,S_v=1,C=2$. Second to fifth rows show simulations of the advanced model (\ref{complexdah}) under the following scenarios: `mixing' ($S_u=S_v=C=8$, second row); `encapsulation' ($S_u=10, S_v=4, C=6$, third row); `partial sorting' ($S_u=10, S_v=8, C=3$, fourth row); `complete sorting' ($S_u=S_v=10, C=0$, fifth row). All other parameters set at $d_u = d_v = R = \kappa_u = \kappa_v = 1$.  The domain $\Omega$ is of size 10$\times$10.}
    \label{adhesionsims}
\end{figure}

As noted earlier, nonlocal formulations can be easily extended to include multiple populations, see (\ref{multiaggregation}). A natural question, therefore, is whether cell sorting can be replicated under a nonlocal formulation. Consider two populations $u$ and $v$ and assume equivalently simple forms to (\ref{kfforms}), then a basic model to describe cell sorting can be stated by the equations
\begin{subequations} \label{simpledah}
    \begin{align}
        \partial_t u & = d_{u} \Delta u -\nabla \cdot \left( u \int_{B^n_R}  \left( S_u u({\bf{x}}+{\bf{r}},t) + C v({\bf{x}}+{\bf{r}},t) \right) \vec{\bf{e}}_r \,d{\bf{r}}\right)\,, \\
        \partial_t v & = d_{v} \Delta v -\nabla \cdot \left( v \int_{B^n_R} \left( S_v v({\bf{x}}+{\bf{r}},t) + C u({\bf{x}}+{\bf{r}},t) \right) \vec{\bf{e}}_r \,d{\bf{r}}\right)\,.
    \end{align}
\end{subequations}
In this model $S_u$, $S_v$ and $C$ represent the $u$-$u$ self-adhesion strength, the $v$-$v$ self-adhesion strength, and the $u$-$v$ cross-adhesion strength, respectively. Note that the interaction ranges are the same (and equal to $R$) and cross interactions are symmetrical, although such assumptions can be relaxed and repelling interactions can also be introduced (for example, see \cite{painter2015,carrillo2018zoology}). Unfortunately, this basic formulation (\ref{simpledah}) proves overly simple to capture the nuances of cell sorting. As for the basic homogeneous model (\ref{simpleadhesion}), the linear choices for the nonlocal terms lead to excessive attraction and the populations become highly concentrated, see Figure \ref{adhesionsims}, top row. The model, as such, is unsatisfactory when it comes to resolving the subtly distinct cell sorting patterns shown in Figure \ref{aggregationsorting}(b). 

Consequently, `successful' nonlocal models \cite{armstrong2006,gerisch2010b,painter2015,murakawa2015,carrillo2019} that are more broadly capable of replicating the spectrum of arrangements predicted by the DAH include modifications to the various terms in model (\ref{simpledah}). For example, this has included adding biologically-meaningful features such as a limitation or saturation to the adhesive pull (see \cite{armstrong2006,gerisch2010b}), introducing volume-filling effects that prevent cell aggregation beyond a critical (packed) level (see \cite{painter2015,carrillo2019}), or modifying diffusion terms to include total population pressure effects (see \cite{murakawa2015,burger2018,carrillo2019}). To provide one concrete example, by adapting the saturating functional forms above and including  population-pressure effects to create sharply segregated boundaries (see \cite{murakawa2015,carrillo2019}), we have 
\begin{subequations} \label{complexdah}
    \begin{align}
        \partial_t u & = \nabla \cdot \left[ d_u u\nabla (u+v) - u \int_{B^n_R} \frac{S_u u({\bf{x}}+{\bf{r}},t)+ C v({\bf{x}}+{\bf{r}},t)}{\kappa_u+u({\bf{x}}+{\bf{r}},t)+ v({\bf{x}}+{\bf{r}},t)}\vec{\bf{e}}_r \,d{\bf{r}}\right]\,, \\
        \partial_t v & = \nabla \cdot \left[ d_v v\nabla (u+v) - v \int_{B^n_R} \frac{S_v v({\bf{x}}+{\bf{r}},t)+ C u({\bf{x}}+{\bf{r}},t)}{\kappa_v+u({\bf{x}}+{\bf{r}},t)+ v({\bf{x}}+{\bf{r}},t)}\vec{\bf{e}}_r \,d{\bf{r}}\right]\,.
    \end{align}
\end{subequations}
This more `advanced' sorting model is capable of replicating the nuances of cellular sorting under different adhesive relationships, e.g. for two populations it can generate the full spectrum of arrangements from mixed to complete sorting see Figure \ref{adhesionsims}.

Summarising, nonlocal models are capable of reaching two touchstones of adhesive behaviour: (i) capturing the adhesive or sticky-like properties of cells in close contact, and (ii) replicating cell-sorting phenomena for heterogeneous adhesive populations as predicted by the DAH. 

At this point we return to our earlier implication that local formulations are incapable of adequately describing adhesion and sorting dynamics, stressing that this applies to `n\"{a}ive' local formulations. In fact, various local models can be shown to exhibit sorting. One method (though not directly describing adhesion) is through extension of a chemotaxis framework: effectively, a `differential chemotaxis' system in which two populations have distinct chemotactic responses to multiple chemical factors (e.g. \cite{painter2009,knutsdottir2014}), so the interactions are indirectly mediated. Directly relevant to adhesion, an intriguing (fourth order) local model has been recently formulated in \cite{falco2023} and demonstrates an impressive capacity to simulate the range of cell sorting patterns described here: we return to this in the discussion.

\subsection{Further applications to cellular systems} \label{s:applications}

Classic cell sorting experiments \cite{townes1955} were first performed using embryonic cell populations, naturally leading to a conjecture that adhesion and sorting are fundamental during embryonic development (for a historial retrospective, see \cite{steinberg2004}). Consequently, a principal application for nonlocal models for cell adhesion lies in developmental processes. In fact, the first nonlocal model for adhesion\cite{sekimura1999} was proposed in the context of self-organisation of scale cells during lepidoptera (moth and butterfly) wing morphogenesis. Nonlocal adhesion models have subsequently been developed, as described above, to show fundamental cell sorting (see \cite{armstrong2006,gerisch2010b,murakawa2015,carrillo2019}), somitogenesis\footnote{A fundamental early embryonic stage of segmented animals, whereby mesoderm tissue is sequentially discretised into blocks of cells along the head to tail axis.}  \cite{armstrong2009}, skeletal morphogenesis\footnote{The embryonic process during which the skeleton is formed.} \cite{glimm2014,bhat2019}, aspects of neural development \cite{matsunaga2017,trush2019}, and vasculogenesis\footnote{Formation of the primitive vasculature network} \cite{villa2022}. Notably, some of these applications have been directly formulated alongside experimental data, linking predictions formed from models to targeted experiments. For example, a nonlocal model of adhesion was formulated\cite{glimm2014,bhat2019} to describe mesenchymal cell movements which indicated a crucial aggregating role for adhesion during early skeletal morphogenesis  .
Experimental-theoretical studies that feature nonlocal adhesion models have also been used to understand brain development, in particular the crucial role of N-cadherin mediated adhesion in the positioning of neuronal populations during mammalian cortex development \cite{matsunaga2017} and the visual centre of the fly {\em Drosophila melanogaster} \cite{trush2019}. 

Abnormal regulation of adhesive processes may be a factor for various pathologies, in particular cancers \cite{janiszewska2020}. For example, a point of significant focus has been on the epithelial-mesenchymal transition (EMT), where upregulation of N-cadherin accompanied by downregulation of E-cadherin allows cells to adopt a more migratory form, linked to increased invasiveness and metastasis \cite{loh2019}. Many mathematical models have been developed to address the roles played by cell-cell (and cell-matrix) adhesion during invasion and a growing number (e.g. \cite{gerisch2008,kim2009,painter2010,domschke2014,bitsouni2017,bitsouni2018,hodgkinson2018,suveges2021}) have applied nonlocal formulations: to understand how adhesion alters the shape of cancer invasion (e.g. \cite{gerisch2008,painter2010}); 
the role of cell-cell adhesion during glioma growth (e.g. \cite{kim2009,suveges2021}); shaping different forms of tumour infiltration patterns in ductal carcinomas (see \cite{domschke2014}); and, two population models, featuring cancer populations at different states of mutation (see \cite{bitsouni2017,bitsouni2018}). Other points of application for nonlocal models of adhesion and cell interactions include wound healing (e.g. \cite{dyson2010,dyson2013,webb2022,webb2023}) and modelling the interactions between liver cells \cite{green2010}.

\section{Nonlocal models for ecological systems}




\subsection{Swarms, flocks, and herds}

Swarming, herding, and flocking phenomena are perhaps the most obvious examples of collective behaviour in ecological systems \cite{sumpter2010collective}. The central idea is that animals, like cells, often exhibit social interactions that cause them to aggregate.  At their most basic level, social interactions may simply cause animals to be found in a particular area of space at some point in time, rather than using all the available area \cite{topaz2006}. At a more advanced level, these interactions can cause a very wide range of complex patterns to emerge, famously exemplified by starling murmurations, but present throughout the animal kingdom \cite{ballerini2008empirical,sumpter2010collective}. 

An enormous number of models have been formulated to understand collective animal movements \cite{vicsek2012,berdahl2018}, a substantial proportion of which are based on systems of `interacting particles'\footnote{In probability theory, the term `interacting particle system' has a specific definition in the context of continuous time Markov jump processes. When we refer to interacting particles within this article, we will often slip into a slightly broader sense: complex systems composed of agents that interact with each other according to their relative positions and/or velocities.}: the position of each agent is governed by a dynamic (usually, stochastic) equation featuring terms that account for how the trajectories of neighbours influence movement (well known models include those in \cite{aoki1982,reynolds1987,heppner1990,vicsek1995novel,couzin2002collective,couzin2005,cucker2007,motsch2011}). Typically, the interactions lying at the heart of these models are formulated according to the `first principles of swarming’ \cite{carrillo2010}. At the shortest range, interactions are often repulsive, as animals will want to avoid physical contact.  At a slightly longer range, animals will align their movements with one another.  Then if animals become too far apart, they have a tendency to move towards one another to maintain the group cohesion (attraction).  These three zones of nonlocal interactions\footnote{One of the earliest and most influential model explicitly built along these principles-- the `Boids' model of Reynolds \cite{reynolds1987} -- was developed with the main aim of generating realistic flocking-like behaviour for the computer graphics industry, rather than the more elementary aim of understanding movement dynamics; numerous interactive online simulators of this model exist, e.g. \texttt{https://boids.cubedhuang.com/}. A particularly notable branch that evolved from that work was the application of swarming models to optimization, i.e. particle swarm optimization \cite{kennedy1995}.} combine to give both stationary and moving aggregations, as well as a vast swathe of spatio-temporal patterns, mimicking many of those that have been observed in nature (see \cite{ sumpter2010collective,vicsek2012,berdahl2018}).

A smaller -- but still substantial -- literature has approached the same central problem of swarming and animal movement via a continuous framework, using ideas that surround nonlocal advection (see \cite{mogilner1999,topaz2006,eftimieetal2007, pottslewis2019, wangsalmaniw2022}).  In fact, the earliest nonlocal biological aggregation models were developed to describe swarming-like behaviour (see \cite{kawasaki1978,mimura1982,nagai1983,mogilner1999,lee2001}) and were based on the nonlocal PDE (\ref{aggregation1}). For example, in \cite{mogilner1999}  even or odd forms of interaction kernels were explored for their capacity to generate drift-type (coherent movement of the swarm) or aggregation-type (cohesion of the swarm) behaviour. A further branch of nonlocal PDE methods are founded on hyperbolic kinetic transport equations (see \cite{eftimieetal2007,eftimie2018hyperbolic,bernardi2021leadership}). In these models, the nonlocal terms do not enter the advection terms, but the turning behaviour of the population; consequently, they benefit from a closer description of individual behaviour and can, for instance, explicitly incorporate the above principles of swarming commonly used in particle models. However, these models represent significant and non-trivial extensions of Equations (\ref{aggregation}-\ref{multiaggregation}) – although it is possible to connect them \cite{buono2015} -- and are more challenging to explore from an analytical and numerical perspective. As such, we do not go into details, instead we refer the reader to a recent book\cite{eftimie2018hyperbolic} that summarises developments in this area.

\subsection{Home ranges and territories via stigmergy and memory}

As well as the visually-impressive examples of collective movement, aggregation phenomena can also occur over longer spatial and temporal scales, becoming apparent as one observes animal locations over a period of time.  For example, by plotting locations over an increasing time window, it often transpires that animals do not use as much of the available area as their locomotive capabilities allow.  Instead they confine themselves to a smaller area called a home range, which they may maintain for a season or even a whole lifetime \cite{burt1943, borgeretal2008}.  This causes the spatial distribution of the animal to tend to a stationary, non-constant distribution, such as can be modelled by Equation (\ref{aggregation2}) or variants thereof \cite{briscoe2002home}.  

Home ranges can emerge due to a range of biological processes.  For example, animals may tend to re-visit locations remembered to be good for foraging \cite{riotte2015memory}.  Once they have memory of sufficiently many locations to meet their foraging needs, they may decide to stay in the vicinity of those locations (see \cite{vanmoorteretal2009, merkle2017energy}).  Additionally, they may need to construct a central place near to where they forage, such as a den or nest site, for reproductive purposes.  The requirement to return to this central place then provides yet another mechanism of locational aggregation \cite{moorcroftetal2006}.  Finally, animals may leave traces of their past locations in the landscape (e.g. through scent marks) and use these as markers to keep them in their home range: a process called stigmergy \cite{theraulaz1999brief}.  In any of these cases, the decisions of the animal to move will tend to be spatially non-local, due to the animals' ability to sense their surroundings as they move, through sight, smell, or memory of target locations (see \cite{potts2014predicting,bastille2018spatial,fagan2020improved}).

To model these biological processes, it is common to couple a nonlocal advection-diffusion equation for the location distribution to an ordinary differential equation (ODE) modelling the process of memory or stigmergy.  The recent review of \cite{wangsalmaniw2022} gives a thorough exposition of these process, but perhaps the simplest example is
\begin{align}
\label{eq:memory1}
\partial_t u & = d\Delta u -\nu \nabla \cdot (u \nabla w_R\ast m), \\
\label{eq:memory2}
\partial_t m &= \alpha u - \delta m,
\end{align}
where $u(\bx,t)$ is the probability distribution of the animal and $m(\bx,t)$ denotes the cognitive map \cite{wangsalmaniw2022}, which models either the density of marks left on the terrain or the amount of memory the animal has about location $\bx$ at time $t$.  Other notation is as in Equation (\ref{aggregation2}). 

Territoriality provides another reason why animals may confine themselves in space over long periods of time.  Here, the presence of neighbouring conspecifics forces animals into a confined space (see \cite{adams2001approaches, potts2014animal}).  There are various mechanisms by which this can happen, but from a modelling perspective they fall into two categories.  The first is via stigmergy: indirect interactions mediated by some form of marks on the terrain, such as urine, faeces, or a trail \cite{moorcroftetal2006, potts2016territorial}.  In this case, animals avoid the marks left by others in the recent past, and usually these marks decay over time.  The second is via memory of direct interactions, such as displays or fights \cite{king1973ecology}.  Animals remember the locations of these displays or fights and may tend to avoid them in the near future \cite{potts2016memory}.  In either case, as with home range formation, the movement of animals in response to these interactions is usually spatially non-local.  

These territorial mechanisms can be modelled using exactly the multi-population system in Equation (\ref{multiaggregation2}) with $\nu_{ij}<0$ for $i \neq j$ to model mutual avoidance, and $\nu_{ii}\geq 0$.  
However, as with home range models, it is often valuable to model the process of memory or stigmergy explicitly via ODEs.  A simple example can be given by combining the ideas behind Equations (\ref{eq:memory1})-(\ref{eq:memory2}) with those of Equation (\ref{multiaggregation2}), as follows 
\begin{align}
\partial_t u_i & = d_i\Delta u_i - \nabla \cdot (u_i \nabla \sum_{j=1}^p \nu_{ij} w_R\ast m_j), \\
\partial_t m_i &= \alpha u_i - \delta m_i,
\end{align}
where $m_i(\bx,t)$ denotes the cognitive map of species $i$, and models the marks left by individuals from territorial unit $i$, whilst $\alpha$ and $\delta$ are constants.  However more complicated versions can be considered that include extra biological realism \cite{potts2016memory, potts2016territorial,wangsalmaniw2022}.

\subsection{A general framework for non-local interactions in ecology}

\begin{figure}[t!]
    \centering
    \includegraphics[width=\textwidth]{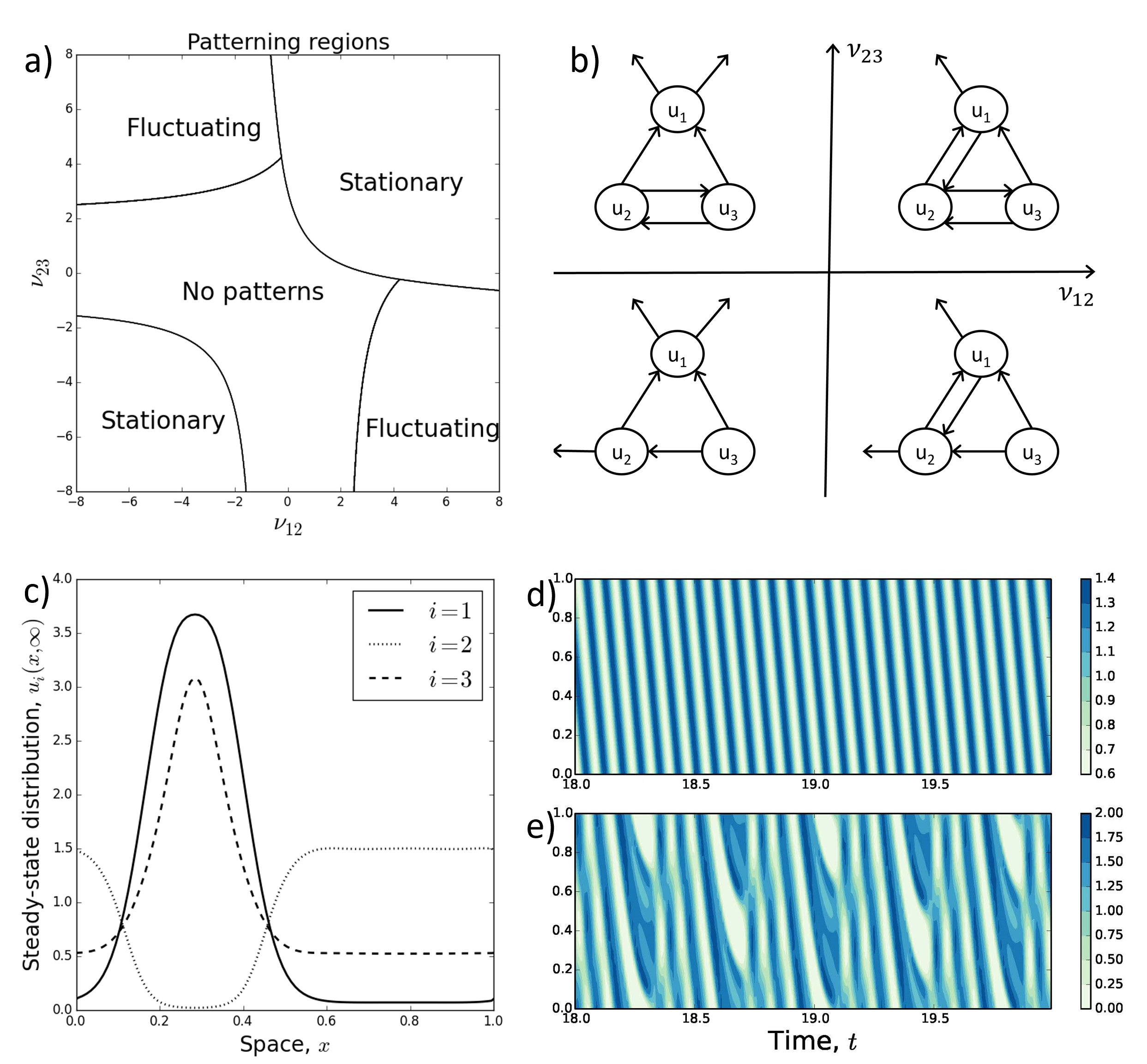}
    \caption{Patterns for example three-species model ecosystems of the form in Equation (\ref{multiaggregation2}).  Panel (a) gives the linear pattern formation regimes for systems described by Panel (b).  In each system, an arrow from $u_i$ to $u_j$ means that $u_i$ is attracted to $u_j$.  An arrow away from $u_i$ in the opposite direction from $u_j$ means $u_i$ avoids $u_j$.  So, for example, the top-left graph in Panel (b) might model two mutualist predator species living alongside a single prey species.  Panels (c-e) give numerical examples of the patterns that can form in a three-species system.  In Panel (c), the system tends to a steady state where $u_1$ and $u_3$ aggregate together but are segregated from $u_2$.  Panels (d) and (e) give example spatio-temporal patterns for $u_1$ with a three-species system.  In all panels, $d_1=d_2=d_3=\nu_{21}=\nu_{31}=\nu_{32}=1$ and $\nu_{13}=-1$.  In Panels (c-e), $\nu_{23}=-4$.  Panels (c-e) have $\nu_{12}=-4$, $\nu_{12}=3.3$, and $\nu_{12}=4$ respectively.}
    \label{ecol_plots}
\end{figure}

As well as territory formation, the multi-species case from Equation (\ref{multiaggregation}) enables a variety of other ecological phenomena to be modelled  over timescales where births and deaths are negligible (e.g. for mammals and birds, this may be over a season or year) \cite{pottslewis2019,giunta2022local}.  For example, the movements of co-existing predators and prey can be modelled by assuming prey advect away from predators and predators towards prey \cite{di2016nonlocal}.  Likewise, competing species may advect away from one another and mutualistic animals may have a tendency to move towards one another. In forager and scrounger interactions, the latter follow the former to exploit their foraging efforts (e.g. see \cite{tania2012}). In ecosystems consisting of many species, there will be a complex network of such interactions that can cause a wide range of emergent patterns (Figure \ref{ecol_plots}c-e).  

As a consequence, Equation (\ref{multiaggregation2}) has been proposed as a key study system for understanding spatial distributions of interacting groups of animals that may emerge over such timescales \cite{pottslewis2019}.  These groups of animals may be territorial groups, populations, or whole species (but we often just use `species' for all such groups for simplicity and consistency with the rest of this review). The overall aim is to be able to provide links between the network of interactions between moving species (Figure \ref{ecol_plots}b) and their pattern formation properties.  

For example, Figure \ref{ecol_plots}a shows the predictions of linear stability analysis for four different systems of three populations (model (\ref{multiaggregation2}) for $i=1,2,3$)   shown schematically in Figure \ref{ecol_plots}b.  This gives a simple categorisation into `no patterns' (all eigenvalues having negative real parts) `stationary patterns' (the dominant eigenvalue is real and positive) or `fluctuating patterns' (the dominant eigenvalue is non-real with positive real part).  However, further away from linear stability regime, patterns in three-population systems can be quite complex and varied, including stationary patterns of aggregation and segregation (Figure \ref{ecol_plots}c), travelling-wave-like solutions (Figure \ref{ecol_plots}d), perpetual irregular oscillations (Figure \ref{ecol_plots}e), and more \cite{pottslewis2019, giunta2022detecting}.

\section{Derivations from the individual level and connecting to data}
\label{sec:deriv}

\subsection{Random walks} \label{s:randomwalks}

When Karl Pearson coined the term `random walk' in 1905 \cite{pearson1905}, the central question involved biological movement: if, within a particular time step, each mosquito moves some distance in a randomly chosen angle, can we estimate the distribution of a mosquito infestation? Fundamental work by Patlak \cite{patlak1953} extended the question to include biases from the environment and persistence. Across the last few decades a vast number of studies have aimed to connect the random walk movements performed by individuals to population level measures and distributions,  for both cell and animal movement (e.g. see \cite{Othmer1988,berg1993,turchin1998,codling2008}). Specifying a position jump random walk (PJRW, see \cite{codling2008,okubo2001,painter2018,stevens1997,Othmer1988}) forms a particularly well trodden path. In the context of the present review, this approach can be used to motivate both local and nonlocal models for aggregation \cite{buttenschon2018}. To illustrate this, we first lay down a general formalism.

Let us consider the probability that a random walker has its centre at position $\bx$ at time $t$. If we have a population of independent walkers, this probability can be equated with the population density $u(\bx,t)$, and we maintain this notion. Note that the definition in terms of the centre implicitly assumes that the walker can have some finite extent, i.e. it is not necessarily a point object. For now we shall avoid any discussion of boundary conditions and assume an individual can move anywhere in space: movement is within $\Omega = \mathbb{R}^n$. The time continuous Master equation for the PJRW has the following form ~\cite{Othmer1988,hughes1995random,van1992}
\begin{equation}\label{Eqn:MasterEquation}
	\partial_t u({\bf x},t) = \lambda \int_{\Omega} 
	[T({\bf x}, {\bf y}) u({\bf y}, t)
	- T({\bf y}, {\bf x}) u({\bf x}, t)]\,
	d{\bf y},
\end{equation}
where $T({\bf x},{\bf y})$ is a probability density function for a jump from ${\bf y} \in \mathbb{R}^n$ to ${\bf x} \in \mathbb{R}^n$. Note that $T$ can depend on $t$, but we omit this dependency from the notation. $\lambda>0$ is a rate parameter. We remark that individuals can remain at their current location through setting $T(\bx,\bx) > 0$, which we refer to as a zero-length jump. We follow the approach of \cite{buttenschon2018} and rewrite the integral kernel $T(\bx,\by)$ according to the jump heading $\bz=\bx-\by$. Specifically,
\[ 
T_\by(\bz) := T(\by+\bz,\by) = T(\bx,\by), \qquad \bz=\bx-\by,
\]
where we assume that 
\[ T_\by\geq 0, \qquad T_\by\in L^1(\R^n), \quad \|T_\by\|_1 =1 .\]
$T_\by$ can be split into even and odd components,
\begin{equation}\label{oddeven1} 
E_\by(\bz) = \frac{1}{2}\left(T_\by(\bz) + T_\by(-\bz)\right), \qquad O_\by(\bz) =\frac{\bz}{2|\bz|} \left(T_\by(\bz)-T_\by(-\bz)\right).
\end{equation}
Then 
\begin{equation}\label{oddeven2}
	T_\by(\bz) =
	\begin{cases}
		E_\by(\bz) + O_y(\bz) \cdot \frac{\bz}{\abs{\bz}} &\mbox{if } \bz \neq 0\\
		E_\by(\bz)                                  &\mbox{if } \bz = 0
	\end{cases}
\end{equation}
with an even part $E_\by\in L^1$ and an odd part $O_\by \in L^1$, which satisfy 
\begin{equation}
	E_\by(\bz) = E_\by(-\bz) \quad \mbox{and} \quad O_\by(\bz) = O_\by(-\bz).
\end{equation}
We employ this decomposition in the general master equation (\ref{Eqn:MasterEquation}) and make two further assumptions. First, that transition rates do not depend on the increment $\bz$, just the starting location $\by$: this describes a myopic random walk. Second, non zero-length jumps are small and of fixed length $h \ll 1$, and Taylor expansions can therefore be applied. Details of the expansions can be found in \cite{buttenschon2018} where, in the limit as $h\to 0$ and $\lambda\to \infty$, we arrive at the advection-diffusion equation 
\begin{equation}\label{Eqn:AdvectionDiffusionLimit}
	\partial_t u(\bx, t) + \nabla \cdot (
	{\bf a}(\bx,t) u(x,t) ) = \Delta ( D({\bf x},t) u({\bf x},t) )\,.
\end{equation}
We denote by $\mathbb{S}^{n-1}$ the $n-1$ dimensional unit sphere in $\R^n$. The advection velocity is given by
\[
{\bf a}(\bx,t) = \lim_{h\to 0,\lambda\to \infty} \frac{\lambda h^n}{n} |\mathbb{S}^{n-1}| \; O_\bx \,, 
\]
and the diffusion term by 
\[ 
D({\bf x},t) = \lim_{h\to 0,\lambda\to\infty } \frac{\lambda h^{n+1}}{2n} |\mathbb{S}^{n-1}|\; E_\bx \,.
\]
Particular care must be paid to the limit scalings, as they suggest different powers of $h$: for the limits to simultaneously exist the odd part must be small (i.e. $O_\bx\sim h$) with respect to the even part. If the odd part is of order one or larger, the diffusion term vanishes and a pure drift equation (a drift-dominated case) is derived. When the odd part is of order $h^2$ or smaller, the drift term vanishes and a diffusion-dominated case arises. The value of separating $T$ with respect to its odd and even parts becomes clear: the even  component $E_\bx$ enters the diffusion term, while the odd component $O_\bx$ determines the advection term. Generally the odd and even parts can involve nonlocal terms that represent sensing up to a certain radius. We will return to this in the next section but one.

\subsubsection{Local models} \label{s:rwlocal}

 We illustrate the above scaling through an interesting local case, which leads to taxis-type models.  To introduce dependency according to some controlling factor, we take the standard assumption\cite{stevens1997} of supposing that the jump probability distribution explicitly depends on a control species, which we denote $c(\bx,t)$. For simplicity, we will restrict in this section to a symmetrical case where we set  $T_\by(\bz) = f(c(\by,t))$ for all non-zero length jumps (i.e. $T$ depends only locally on $\by$ through $f(c(\by,t))$). When movement occurs, all headings are chosen with equal probability, but this probability varies with the local level of the control species $c$. There is no odd component to $T$ and the limiting equation (\ref{Eqn:AdvectionDiffusionLimit})  in this case is of the form
\begin{equation}\label{myopiclocal}
	\partial_t u = d \Delta ( f(c) u ) = d \nabla \cdot \left[ f(c) \nabla u + u f^{\prime}(c) \nabla c \right]\,.
\end{equation}
Therefore -- despite an absence of directionality to the jump -- a taxis-like process emerges at the macroscopic level: advection according to the gradient of $c$. The control species can be distinctly interpreted according to the movement process. For example, it may simply define a fixed environmental variability, e.g. regions where movement is easier or more difficult. It could also change according to the distribution of the population -- for example, a scent deposited by an animal or a chemical released by a cell -- and therefore defined by an evolution equation such as (\ref{eq:memory2}). We refer to \cite{hillen2009user,bellomo2015} for detailed reviews on chemotaxis models.

Using cell adhesion as a case study, a simple but na{\"{i}}ve approach would be to directly equate the control species with the population density. Specifically, we consider $c \equiv u$ and hence obtain the density-dependent diffusion equation
\begin{equation}\label{eq:naive}
\partial_t u = \nabla \left[ D(u) \nabla u \right] \quad \mbox{with} \quad D(u) = d \left( f(u)+ u f^{\prime} (u) \right)\,.
\end{equation}
Considering the `stickiness' property of adhesion, a logical choice for $f(u)$ would be a decreasing function that reflects reduced capacity to move as a cell forms adhesive attachments with its neighbours. For example, a choice $f(u) = \frac{1}{\kappa + u}$ results in $D(u) = \frac{d\kappa}{(\kappa+u)^2}$: this reduces diffusivity in regions of higher population density, and corresponds with certain choices\cite{hofer1995} in macroscopic (phenomenological) approaches to adhesion. 

Derivations of local models for adhesion that rely on the PJRW framework have been considered previously (e.g. see \cite{anguige2009,johnston2012,johnston2013}). 
While more sophisticated than the above -- for example, more complicated jump probabilities or accounting for correlations in movement -- they essentially lead to the same result of a density-dependent diffusion equation.
Clear advantages lie in that they can lead to models that can be fitted against experimental data (e.g. obtained from cell assays \cite{johnston2014,johnston2014b}), and that the derived PDE form is relatively tractable, both analytically and numerically. 

However, while density-dependent diffusion captures one expected consequence of adhesion, it is more questionable in the context of self-organisation or cell sorting phenomena. The possibility of biological aggregation within both the underlying discrete master equation and its corresponding continuous model has been considered in various studies (for example see \cite{lizana1999,painter2003,horstmann2004,anguige2009}), and for (\ref{eq:naive}) it is straightforward to use linear stability analysis (see Section \ref{s:linearstability}) to show that for (\ref{eq:naive}) this will depend on the shape of $f(u)$: instability of the uniform steady state, and hence self-organising capacity, requires $f(u)+ u f^{\prime} (u) <0$. This is not possible for $f(u) = \frac{1}{\kappa + u}$, but can be satisfied when $f(u) = \frac{1}{(\kappa + u)^q}$ for $q>1$. However, at this point the PDE (\ref{eq:naive}) will become illposed and unpractical for application.

\subsubsection{Nonlocal models}

Intuitively, it is the pointwise nature of the dynamics that proves problematic in the above. The random walker responded only to the strictly local information acquired at its centre: it is a point particle, and 
the population can potentially become trapped at singular locations of `infinite stickiness'. 

A cell or organism, though, has a spatial extent and, even if interacting only through direct contact, will interact across some volume of space. This naturally leads to the question of how one can extend derivations from PJRWs in a manner that retains this nonlocality. We will again use cell adhesion as a case study and follow the approach in \cite{buttenschon2018}. As noted earlier, the formation of adhesion bonds between membranes leads to the generation of (local) forces that draw cells together; cellular membranes are highly dynamic, extending and retracting protrusions that span shorter (e.g. lamellipodia) and longer (e.g. filopodia) ranges. Adhesive attachments, therefore, can create forces at a position ${\bx+\br}$ that act to displace a cell centred at $\bx$ where the distance $\br$ is potentially several mean cell diameters away. The method in  \cite{buttenschon2018} is to consider a biased random walk where the bias results from summing over all possible local forces that can impact on the cell centred at $\bx$, which enter the odd component of $T$ in (\ref{oddeven2}). Following the scaling, one obtains a nonlocal advection velocity of the form
\begin{equation}\label{generala}
{\bf{a}}(\bx) = \underbrace{\mu}_{\mbox{\tiny adhesive strength}} \int \underbrace{N_b(u(\bx+\br,t))}_{\mbox{\tiny number of bonds}}  \; \underbrace{S(u(\bx+\br,t))}_{\mbox{\tiny free space}}  \; \underbrace{\omega(|\br|)}_{\mbox{\tiny cell extension}}
\underbrace{\vec{\bf{e}}_r}_{\mbox{\tiny direction}}\, d\br\,. \end{equation}
In the above, $\mu$ denotes an adhesive strength per adhesion bond, $\br$ denotes the direction and length of the cell extension, $N_b(u(\bx+\br,t))$ denotes the bound adhesion receptors that are generated with cells at location $\bx+\br$, $S(u(\bx+\br,t))$ indicates the amount of free space available for cells to extend into this area, $\omega(|\br|)$ denotes the ability of a cell to express adhesion receptors a distance $|\br|$ away from its centre, and $\vec{\bf{e}}_r$ accounts for that bonds generated at $\bx+\br$ will lead to a bias corresponding to that direction.

The formulation in (\ref{generala}) is rather general, therefore admitting varying degrees of biological detail. For example, assuming compact support for the cell extension, no space limitation ($S=1$), and using mass action kinetics to set the number of bonds to be proportional to the cell density ($N_b(u) \propto u$), one essentially arrives at a model of the form (\ref{simpleadhesion}). If, rather, one takes the adhesion binding to be governed by a Michaelis-Menten type binding mechanism, $N_b(u(\bx)) \propto \frac{u(\bx)}{\kappa + u(\bx)}$, then we arrive at a model similar to that specified in (\ref{saturatingadhesion}).

Consequently, through an explicit derivation from a PJRW it is possible to motivate and clarify the implicit assumptions that underlie various nonlocal models for adhesion, in particular those originally developed with phenomenological reasoning and applied to various phenomena (Section \ref{s:applications}). More generally, given that the integral (\ref{generala}) will typically be a nonlinear function of the cell density $u(\bx+\br)$ and the ability to form attachments varies with the distance from the cell centre, one can straightforwardly obtain the general formulation in (\ref{aggregation1}).

\subsubsection{Step selection functions: connecting to data on organism movement}
\label{sec:ssa}

The formalism of a PJRW also allows for relatively straightforward parameterisation of advection-diffusion equations based on data, an approach that has been used both for experimental data obtained for cell systems (say, using cellular assays, e.g.\cite{johnston2014b}) and locational data for animals (e.g. \cite{potts2020parametrizing,potts2023scale}). 

Taking the example of animal movement, these data typically arrive as a time series of locations. If this time series is relatively low frequency, e.g. of the order of one location every few minutes or hours, we might use the funtion $T(\bx,\by)$ 
(Equation \ref{Eqn:MasterEquation}) to model movement between successive measured locations, from $\by$ to $\bx$ (see \cite{forester2009accounting}). Alternatively, if the time series is very high frequency, e.g. many locations per second, which is increasingly common \cite{williams2020optimizing}, $T(\bx,\by)$ can be used to model movements between successive places where the animal makes a turn \cite{munden2021did}.  In this latter case, we are more accurately modelling behavioural decisions of animals, as they will likely turn for a reason \cite{wilson2013turn}.

Either way, a huge amount of ecological insight has been gained in recent years by fitting functions that describe a position-jump process to time series of animal location data (e.g. see \cite{fortin2005wolves,thurfjell2014applications,fieberg2021guide}).  Moreover, further understanding can be gained by scaling these processes up to distributions of broad-scale space use patterns via advection-diffusion equations, using similar techniques to those described in Section \ref{s:randomwalks} \cite{potts2023scale}.  The specific position-jump model that has gained particular interest from the ecological community goes under the name `step selection function' (SSF) and has the following form\footnotemark\footnotetext{The nomenclature in the literature is not always consistent here.  Sometimes SSF refers to Equation (\ref{eq:ssf}), sometimes to the numerator of this equation, and sometimes just to the function $w(\bx,\by)$.}
\begin{align}
\label{eq:ssf}
T(\bx,\by)=\frac{\psi(\bx,\by)w(\bx,\by)}{\int_\Omega \psi(\bx,\by)w(\bx,\by)d\bx},
\end{align}
where $\psi(\bx,\by)$ represents something about the organism's movement capability, often a distribution of `step lengths' $|\bx-\by|$\footnotemark\footnotetext{More generally, $\psi$ could be a distribution of step lengths and turning angles, so that $\psi$ is dependent upon $\bx$, $\by$, and also the bearing $\theta$ on which the animal travelled to $\by$.  But to keep the exposition simple, we will assume here that $\psi$ only depends upon $\bx$ and $\by$.
}, and $w(\bx,\by)$ is a `weighting function' which encapsulates anything that covaries with movement. 
Typically, $w(\bx,\by)$ is written in the following exponential form
\begin{align}
\label{eq:weight}
w(\bx,\by)=\exp[\boldsymbol{\beta}\cdot\mathbf{Z}(\bx,\by)],
\end{align}
where $\mathbf{Z}(\bx,\by)$ is a vector of functions, each of which represents a movement covariate, and $\boldsymbol{\beta}$ is a vector denoting the relative contribution of the effect of each covariate on movement.  In many practical examples of step selection, $\mathbf{Z}(\bx,\by)$ are simply static environmental features measured at the end point of the step (so $\mathbf{Z}(\bx,\by)$ can be written as $\mathbf{Z}(\bx)$) (see \cite{thurfjell2014applications,fieberg2021guide}).  However, they can also represent features along a step, such as barriers \cite{beyer2016you}, or   dynamic quantities such as memory \cite{merkle2014memory} or the presence of other organisms \cite{potts2022assessing}.  Memory processes lead to self-interaction, which may give rise to a single species aggregation-type equation (\ref{aggregation}).  If co-moving animals or interacting populations are being modelled, it is necessary to write a different step selection function for each entity (individual or population) \cite{potts2014unifying}.  These coupled step selection functions then lead to multi-species equations, like Equation (\ref{multiaggregation}) \cite{potts2020parametrizing}.

A reason for the popularity of the functional form in Equations (\ref{eq:ssf}-\ref{eq:weight}) is that parametrisation can be done simply and quickly using conditional logistic regression.  Details of this technique are given elsewhere (see \cite{forester2009accounting,avgar2016integrated}), but in short it involves first approximating the integral in the denominator of Equation (\ref{eq:ssf}) by sampling from $\psi$, and then recognising the resulting function as the likelihood of a case-control study where the samples are the controls.

Although there are many empirical studies using step selection functions to infer information about animal movement (e.g. see \cite{thurfjell2014applications,fieberg2021guide}), there are far fewer that take the next step of deriving the associated advection-diffusion equation to understand broad-scale space use patterns \cite{potts2023scale}.  Perhaps the reason for this is that such studies combine empirically-driven questions with relatively-advanced mathematical analysis, thus require strong interdiscplinary collaborations between applied mathematicians, empirical ecologists, and statisticians.  The flip-side is that there is huge, fertile ground for mathematicians to collaborate with those ecologists involved in step selection studies, enhancing their data analysis and answering new scientific questions \cite{potts2022assessing}.

\subsection{Derivations from interacting particle system models}

As mentioned earlier, many of the ABM-based approaches to cellular and animal aggregation phenomena fall into the broad class of systems of `interacting particles'. Deriving continuous models from these models forms a very large field, and a growing literature has emerged in which nonlocal models related to (\ref{aggregation}) are obtained. It is significantly beyond the scope of the present article to provide a comprehensive examination of this literature. Rather, we provide a few apposite examples and refer to others (e.g. \cite{carrillo2010,motsch2014}) for a more general review.

To provide some context, we consider the following concrete example\footnote{We note that this particular example comes from a model formulated for opinion dynamics, rather than biological aggregation. However, the underlying principles are the same: a tendency to converge, whether in position or opinion, when agents are sufficiently close.} in one dimension; we refer to \cite{motsch2014,garnier2017} for more details. Let the position $x_i(t)$ of agent $i$ in a population of size $N$ at time $t$ be determined by the stochastic differential equation
\begin{equation}\label{IPS}
dx_i = -\frac{1}{N} \sum_{j=1}^{N} a_{ij} (x_i-x_j) dt + \sigma dW_{i} (t)\,.
\end{equation}
In (\ref{IPS}), the $W_i$'s denote independent Brownian motions and model an uncertainty to the particle position (with strength $\sigma$). Interactions are incorporated through the summed term, where $a_{ij}$ gives the strength of interaction between agents $i$ and $j$; the $1/N$ factor averages across all possible interactions. This general form can be tailored to describe an attraction process between sufficiently close individuals -- e.g. as relevant for cell adhesion -- by setting the interaction to be a function of the distance of separation, $\left| x_i-x_j \right|$, with compact support: i.e. no attraction above a critical interaction range. 

To obtain a continuous model, one can consider the following empirical probability measure for the positions of all agents at time $t$: 
\[
u^{N} (t)= \frac{1}{N} \sum_{i=1}^{N} \delta_{x_i(t)} (dx)\,,
\]
where $\delta_x(dx)$ is the Dirac measure with point mass at position $x$. Through application of mean field asymptotic theory, it can be shown \cite{garnier2017} that as $N \rightarrow \infty$ the probability measure $u^{N}$ (weakly) converges to a deterministic density $u$, which under certain conditions is governed by a nonlocal PDE of the form (\ref{aggregation1}).

A number of other derivations from IPS models have also led to equations related to (\ref{aggregation}). In one paper\cite{middleton2014} the starting point was an off-lattice centre-based model (see Section \ref{s:discreteadhesion}), in which the motion of each particle is governed by Newton’s second law of motion under viscous forces, forces from self-propulsion and forces from interactions. The latter allowed adhesion-type interactions to be included, which followed the standard assumption of varying with the degree of separation. A hierarchical system of $N$ nonlocal PDEs was obtained to describe the distribution of a population of $N$ interacting cells and, again following a mean field approximation, a nonlocal aggregation model of the form (\ref{aggregation1}) is obtained.  

Nonlocal aggregation models of the form (\ref{aggregation2}) can also be motivated from an IPS (e.g. see \cite{morale2005,burger2007}). The motivation in \cite{morale2005} lay in the aggregating tendency of ants ({\em Polyergus rufescens}), with each ant's position evolving according to a stochastic differential equation driven by Brownian motion and an interaction drift; drift dominated over random wandering when other individuals enters an ant’s interaction range. Both aggregating and repelling effects were included, with the former operating when another individual enters an attracting range and a repulsion term for when they become too close. Assuming a large population $N$, then in the limit $N\rightarrow \infty$ the following equation was derived for the population density:
\[
\partial_t u = d \Delta u + \nabla \cdot \left[ u \nabla u -u \nabla \left( w \ast u \right) \right]\,,
\] 
where $d$ follows from the Brownian motion and the density-dependent (degenerate diffusion) and nonlocal drift terms follow from the repulsion-attraction interaction; $w$ derives from the aggregation interaction kernel. The above essentially combines the formulation (\ref{aggregation2}) with an additional degenerate diffusion term, as previously described in Section \ref{s:aggregation}.

\color{black}

\section{Analytical properties}

\subsection{Linear stability analyses} \label{s:linearstability}

A linear stability analysis can be used to demonstrate basic criteria for aggregation from a dispersed initial state, i.e. self-organisation properties. We first consider the formulation (\ref{aggregation1}) and, for maximum clarity, utilise the simple assumptions that lead to (\ref{simpleadhesion}) and constrain to a one-dimensional infinite domain; the latter restriction circumvents the complications that arise from specific boundary conditions.  Consequently, (\ref{simpleadhesion}) becomes
\begin{equation} \label{simple1_1d}
\partial_t u = d\partial_{xx} u - \mu(R) \partial_x \left[ u \left(\int_0^R u(x+y,t) dy - \int_0^R u(x-y,t) dy \right) \right] \,,
\end{equation}
while under equivalent assumptions the formulation (\ref{aggregation2}) becomes
\begin{equation}\label{simple2_1d}
\partial_t u = d\partial_{xx} u - \nu(R) \partial_x \left[ u \partial_x\left(\int_{-R}^{R} u(x+y,t) dy \right) \right] \,.
\end{equation}
Assuming that the population is initially distributed about a uniform steady state $\bar{u}$, we perform a Turing-type stability analysis (e.g. \cite{murray2003}) by linearising about the uniform steady state and looking for solutions to the linearised equation with mode $k$ and eigenvalue $\sigma$ as as $e^{ikx+\sigma t}$. This yields the characteristic equations for the eigenvalue-wavenumber relationship
\begin{equation}
\sigma = - d k^2 + 2 \bar{u} \mu(R) (1 -\cos(kR)) \quad \mbox{and} \quad
\sigma = - d k^2 + 2 \bar{u} \nu(R) k \sin(kR)
\end{equation}
for (\ref{simple1_1d}) and (\ref{simple2_1d}), respectively.  Inhomogeneous perturbations of the steady state grow if there are unstable wavenumbers $k\ne 0$, i.e. those for which $\Re (\sigma(k))>0$. Straightforward inspection of the above reveals that this will hinge on the competition between stabilising (diffusion) and destabilising (aggregation) processes. In particular, the parameter regions in which self-organisation occurs\footnote{Note that this is under the infinite domain assumption, thereby allowing patterns to grow with unbounded wavelengths.} are given by
\begin{equation} \label{patterningspace}
\bar{u}\mu(R) R^2  > d \quad \mbox{and} \quad
2 \bar{u} \nu(R) R  > d
\end{equation}
for the formulations (\ref{simple1_1d}) and (\ref{simple2_1d}), respectively.
While phenomenologically similar, these two conditions are subtly distinct according to the relationships between the strength and range parameters. 

\begin{figure}[t!]
    \centering
    \includegraphics[width=\textwidth]{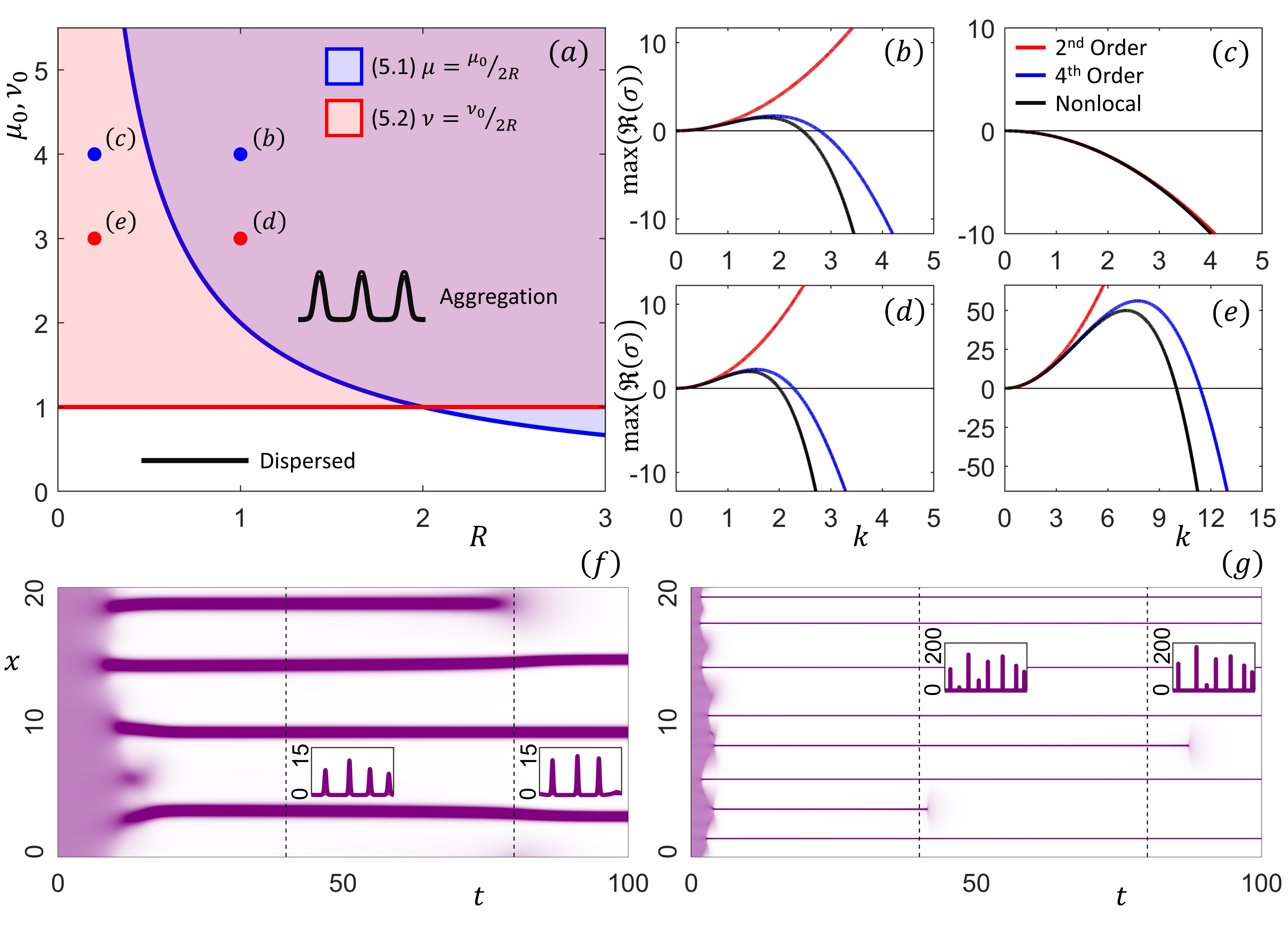}
    \caption{(a) Parameter spaces for self-organisation as predicted by linear stability analysis, for (\ref{simple1_1d}) and (\ref{simple2_1d}) under $\mu(R) = \mu_0/2R$ and $\nu(R) = \nu_0/2R$, respectively. (b-e) Representative curves for the characteristic equations, corresponding to the points highlighted in (a): (b-c) formulation (\ref{simple1_1d}) and its second and fourth order approximations; (d-e) (\ref{simple2_1d}) and its second and fourth order approximations. (f-g) Simulations of (\ref{simple1_1d}) in 1D, for: (f) $(\alpha,R) = (1,3)$; (g) $(\alpha,R) = (0.1,21)$ ; density maps show the population density (white = low density, purple = density $\ge 4 \bar{u}$), with inset figures showing the profile at the two times indicated by the dashed lines. For all plots, other parameters are set at $d = \bar{u} = 1$.}\label{fig:linearstability}
\end{figure}

Commonly, the nonlocal terms in models of type (\ref{aggregation}) are normalised, e.g. according to a measure of the size of the interaction space: in the context of (\ref{simple1_1d}-\ref{simple2_1d}), it is standard to choose $\mu(R)  = \mu_0/2R$ and $\nu(R)  = \nu_0/2R$. Under this choice, the instability conditions for the interaction strength ($\mu_0$ or $\nu_0$) and interaction range ($R$) have some clear distinctions for the two models (\ref{simple1_1d}-\ref{simple2_1d}) and become
\[
\bar{u} \mu_0 R   > 2 d \quad \mbox{and} \quad
 \bar{u} \nu_0  > d, \quad \mbox{ respectively.}
\]
The condition for (\ref{simple2_1d}) is independent of $R$, while for (\ref{simple1_1d}) the capacity for self-organisation is lost as the interaction range decreases. We illustrate the parameter spaces in Figure \ref{fig:linearstability}(a).

Characteristic equation curves for particular parameter values illustrate these behaviours: large $R$ and sufficient $\mu_0$ or $\nu_0$ allows patterning for both models; correspondingly, a finite range of unstable wavenumbers is observed (Figure \ref{fig:linearstability}(b,c), black curves). Decreasing $R$, the range of unstable wavenumbers either expands for (\ref{simple2_1d}) (Figure \ref{fig:linearstability}(e)) or shrinks and disappears for (\ref{simple2_1d}) (Figure \ref{fig:linearstability}(d)).

Further insights are obtained through expanding $u(x\pm y)$ inside (\ref{simple1_1d}-\ref{simple2_1d}) and truncating at different orders. The simplest nontrivial case (using $\mu = \mu_0/2R$, $\nu = \nu_0/2R$) leads to the second order local approximations 
\begin{equation} \label{secondorder}
\partial_t u= \partial_x \left[ \left(d - \frac{\mu_0 R}{2} u \right) \partial_x u \right] \quad \mbox{and} \quad \partial_t u = \partial_x \left[ (d - \nu_0 u ) \partial_x u \right]
\end{equation}
for (\ref{simple1_1d}) and (\ref{simple2_1d}), respectively. These immediately recall the density dependent diffusion forms derived in Section \ref{s:rwlocal}. Instability regions for these equations are identical to those defined by (\ref{patterningspace}), however this coincides with the region in which the models become illposed (negative diffusion); this manifests through corresponding  characteristic equations whereby all wavenumbers are unstable, see red curves in Figure \ref{fig:linearstability} (b-e). 

The expansions can also be truncated at higher order terms, and in particular the fourth order approximations become
\begin{equation} \label{fourthordera}
\partial_t u = \left[ \left(d - \frac{\mu_0 R}{2} u \right) \partial_x u - \frac{\mu_0 R^3}{48} u \partial_{xxx} u \right]
\end{equation}
and 
\begin{equation} \label{fourthorderb}
 \partial_t u = \partial_x \left[ \left(d - \nu_0 u \right) \partial_x u - \frac{\nu_0 R^2}{6} u \partial_{xxx}\right]
\end{equation}
for (\ref{simple1_1d}) and (\ref{simple2_1d}), respectively. Instability regions are again as those defined by (\ref{patterningspace}). However, we now note that the destabilising second order term is countered by a stabilising fourth order term. The characteristic equations in this case generate finite ranges for unstable wavenumbers (blue curves, Figure \ref{fig:linearstability} (b-e)), curves closely following those of the nonlocal model (black curves). The distinct limiting behaviours as $R\rightarrow 0$ become clear from (\ref{fourthordera})-(\ref{fourthorderb}): the fourth order approximation to (\ref{simple1_1d}) implies convergence to a simple diffusion equation, with constant (and nonnegative) diffusion coefficient $d$; the fourth order 
approximation to (\ref{simple2_1d}), however, converges to a density dependent form (second equation in (\ref{secondorder})) with potential illposedness. We note that the fourth order approximations to nonlocal models have been studied in detail, e.g. in \cite{sekimura1999} for one variable models and in \cite{falco2023} for two variable models (see also Discussion and Challenges). 

Stability analyses can, of course, be extended to explore pattern formation in multi-species models, for example those formulated to simulate adhesion-driven cell sorting. Scenarios under which pattern formation can occur will inevitably become more complicated within such models, as there are more potential routes to pattern formation (e.g. through the self interactions or through the cross interactions). We refer to \cite{painter2015,pottslewis2019} for an examples of stability analyses for multi-species  situations.

\subsection{Global existence and boundedness} \label{s:existence}

Our above observation of illposed local models that can follow from approximations of (\ref{aggregation}) leads to questions regarding the local and global existence of solutions: numerical solutions suggest that aggregates can become highly concentrated (e.g. Figure \ref{adhesionsims1}(b)), but still appear to approach a bounded form. Does the presence of the nonlocal term lead to existence of solutions? This has formed a key point of inquiry for a number of publications (e.g. see \cite{laurent2007,sherratt2009, bertozzi2010,dyson2010,chaplain2011, fetecau2011,hillen2018}) related to (\ref{aggregation}).  

For (\ref{aggregation1}), perhaps the most general theory\cite{hillen2018} considers the following form of system
\begin{equation}\label{ArmstrongNd}
\partial_t u = d \Delta u - \mu \nabla \cdot \left( u \int_{B_R (\bx)} f (u (\bx+\br,t) ) \frac{\br}{|\br|} \omega(|\br|) d \br \right),
\end{equation}
where $B_R (\bx)$ denotes the ball of radius $R >0$ around $\bx$. 

\begin{theorem}[Corollary~2.4 in~\cite{hillen2018}]\label{t:Winkler} 
    Assume:
    \begin{enumerate}[label=\textbf{(A\arabic*)},ref=(A\arabic*),leftmargin=*,labelindent=\parindent]
    \item\label{Existence:Assumption:1} $f\in C^2(\R^n)$ and there exists a value $b>0$ such that $f(u)=0$ for all $u\geq b$;
    \item\label{Existence:Assumption:2} $\omega\in L^1(\R^n)$, $\omega\geq 0$;
    \item\label{Existence:Assumption:3} for $p\geq 1$ let $u_0\in X_p:= C^0(\R^n) \cap L^\infty(\R^n)\cap L^p(\R^n)$ be non-negative.
\end{enumerate}
Then there exists a unique, global solution
    \[
	u\in C^0([0, \infty); X_p) \cap C^{2,1}(\R^n\times (0,\infty))
    \]
   of~\eqref{ArmstrongNd} in the classical sense, with $u(\bx,0) = u_0(\bx)$, $\bx\in \R^n$.
\end{theorem}
We remark that while the above immediately implies global existence of solutions in $n$-dimensions for a large class of formulations, it does not yet cover some standard choices. The oft-used formulation (\ref{simpleadhesion}) is particularly delicate as, formally, \ref{Existence:Assumption:1} states that $f$ can only be linear up to a bounded density, but then becomes zero beyond some higher density. From the point of practical application this is sufficient, as we would naturally expect a bound to arise from physical or biological constraints, e.g. space limitations or saturation of receptors. Nevertheless, covering the case $f(u)=u$ without that explicit assumption remains an open problem. 

The same Theorem \ref{t:Winkler} can also be used in the context of other aggregation models, and in particular we refer to the
formulation based on energy minimisation, (\ref{aggregation2alt}). 
To see this, we first note the connection of (\ref{ArmstrongNd}) to the energy-based formulation by supposing there exists some $W(|\br|)$ such that $\nabla W(|\br|) = \frac{\br}{|\br|}\omega(|\br|)$. Recalling that $\br = \by - \bx$, straightforward calculations (shown in \ref{a:integral}) reveal that (\ref{ArmstrongNd}) can be rewritten as 
\begin{equation}\label{energymin}
\partial_t u = d \Delta u +\mu \nabla \cdot \left( u \nabla (W \ast f(u)) \right)\,.
\end{equation}
Therefore, we can apply Theorem \ref{t:Winkler} straightforwardly:
\begin{corollary}\label{Cor:carrillo}
Consider the model (\ref{energymin}) where $\mu>0$ and $W(|\br|)$ is a potential, a function of the distance of the interaction $\left|\br \right| = \left| \by-\bx \right|$. Suppose $f$ satisfies the same conditions as (A.1), the initial condition satisfies (A.3) and $W$ satisfies 
\begin{enumerate}[label=\textbf{(W\arabic*)},ref=(A\arabic*),leftmargin=*,labelindent=\parindent]
    \item $W(|\br|)\in L^\infty$, and 
     $W(|\br|)$ has compact support inside a ball $B_R (0)$. 
    \item There exists a scalar function $\omega(|\br|)$ such that $\nabla W (|\br|)=  \frac{\br}{|\br|}\omega(|\br|)$ 
    where $\omega(|\br|) \in L^1$ and $\omega(|\br|)\geq 0$. 
\end{enumerate}
Then equation (\ref{energymin}) has a unique global classical solution 
 \[
	u\in C^0([0, \infty); X_p) \cap C^{2,1}(\R^n\times (0,\infty)).
    \]
\end{corollary}
{\bf Proof.} The proof
follows immediately from Theorem \ref{t:Winkler} by replacing $\nabla W$ with $\frac{\br}{|\br|} \omega(|\br|)$. 

\begin{figure}
    \centering
\includegraphics[width=\textwidth]{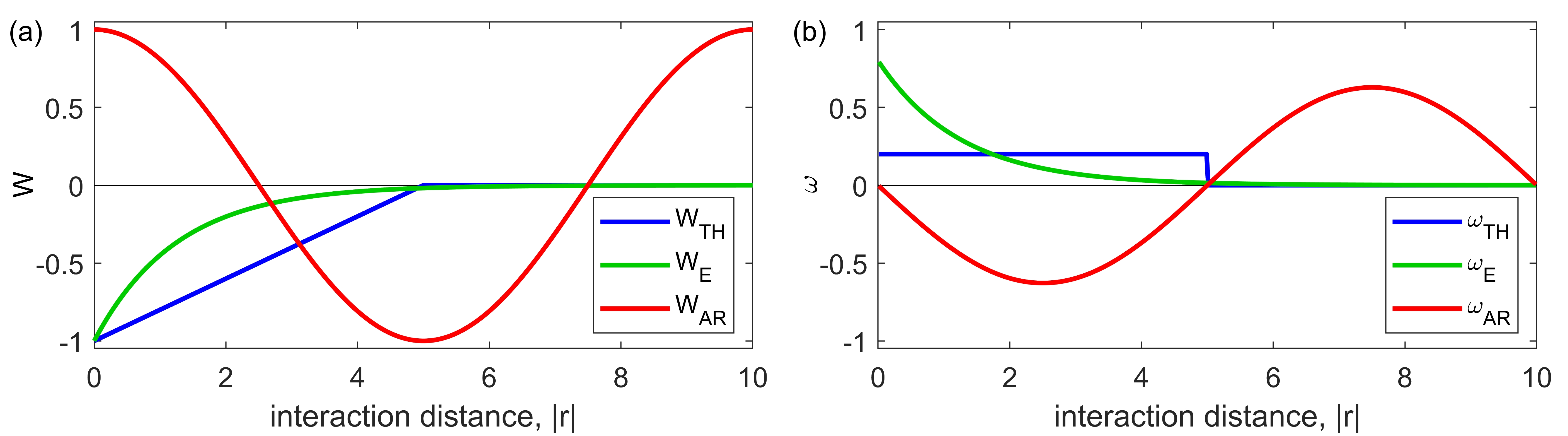}
    \caption{Examples of interaction potentials (left) and the corresponding forces (right), using $R=5$. Here we consider the top hat potential $W_{TH}$, the exponential potential $W_E$, and the attraction-repulation potential $W_{AR}$.} \label{f:potentials}
\end{figure}

Corollary \ref{Cor:carrillo} is the first existence result for aggregation models (\ref{aggregation2}) with general nonlinear response functions $f(u)$. However, the condition (W3) is quite restrictive. Since we require $\omega(|\br|)\geq 0$, condition (W2) imposes that the drift is always towards the origin, where the origin corresponds to the location of the probing individual. Hence the forces are always attractive. Examples of attractive potentials are shown in Figure \ref{f:potentials}A, and include the linear potential $W_{TH}$ and the exponential potential (also called a Moore potential or Laplace kernel) $W_E$, 
\[ W_{TH}(|\br|) = \min\left\{\frac{1}{R} |\br| -1,0\right\}, \qquad W_E(|\br|) = - \exp\left(-\frac{4 |\br|}{R}\right), \]
where $R$ represents an interaction range parameter. The exponential kernel has unbounded support, but converges to zero quickly for larger $|\br|$; the factor of four ensures that this is close to zero for $|\br|=R$. Other purely attractive potentials include the Gaussian kernel and the Hegselman-Krause potential used in opinion dynamics (e.g. see \cite{Lutscherbook,giunta2022local,LPP16,Hegselman}). 

In the cases described above the potential is strictly increasing for small values of $|\br|$, hence indicating an attractive force towards the origin. Indeed, the corresponding kernels $\omega(|\br|)$ are nonnegative, see Figure \ref{f:potentials} (b). As a point of note, under the linear potential $W_{TH}$ we obtain a so-called top-hat kernel, e.g. as previously used in (\ref{simpleadhesion}).

In the swarming literature it is quite common to consider potentials that describe both, attractive and repulsive effects. In such cases $W$ is no longer monotonic, and hence $\omega$ changes sign: when $W$ is increasing, $\omega>0$, and we are in an attracting region; if $W$ is decreasing, $\omega<0$, and we are in a repelling region. One simple example of an attraction-repulsion potential, also shown in Figure \ref{f:potentials}, is given by
\[W_{AR}(|\br|) = \cos\left(\frac{\pi |\br|}{R}\right)\,.\]
This stipulates a repelling region for interaction distances up to $R$, and an attracting region from $R$ to $2R$. Note that the attraction-repulsion potential has a minimum at $|\br|=R$, at the point at which there is a switch from repulsion to attraction, and in this context $R$ can be regarded as the preferred distance between individuals. Other examples of attractive and repulsive potentials are discussed in \cite{Carrillo2020} and include the generalized Kuramoto model, the Onsager model for liquid crystals, and the Barr\'e-Degond-Zatorska model. 

\subsection{Bifurcation analysis} \label{s:bifurcation}

There are two principal techniques that have been used to analyse bifurcations in models of the type in Equations (\ref{aggregation}) and (\ref{multiaggregation}): weakly nonlinear analysis (WNLA) and Crandall-Rabinowitz bifurction theory (CRBT). Both techniques are useful for separating bifurcations into sub- and super-critical regimes.  However, CRBT relies on steady-state formulations, whereas WNLA can reveal the criticality of bifurcations whereby the dominant eigenvalue is non-real and so solutions just beyond the bifurcation point oscillate in time.  On the other hand CRBT can be used to understand the global nature of branches \cite{hillen2021}, whereas WNLA is intrinsically local in its formulation \cite{MC00}.  We give examples here of both techniques, first CRBT then WNLA, applied to our models of interest, exemplifying valuable outcomes and important considerations when applying them.

\subsubsection{Crandall-Rabinowitz type bifurcation analysis} 

Bifurcation analyses that uses the Crandall and Rabinowitz framework \cite{Crandall1971,Rabinowitz1971} (alongside methods from equivariant bifurcation theory \cite{golubitsky2003}), have been carried out in a recent monograph
\cite{hillen2021}. To illustrate, we consider a particularly simple setting in one dimension, for the interval domain $[0,L]$ with a possibly non-linear adhesion function $f(u)$:
\begin{equation}\label{bif1}
\partial_t u =  \partial_{xx} u -\mu \partial_x \left[u \int_{-1}^1 f(u(x+r,t)) \frac{r}{|r|} \omega(|r|) dr\right], 
\end{equation}
where $\omega(|r|) \geq 0$, $\omega\in L^1(0,1)\cap L^\infty(0,1)$, and $\|\omega\|_{L^1(0,1)} = \frac{1}{2}$. In (\ref{bif1}) we implicitly assume that the integral kernel has compact support in the interval $[-1,1]$ and that $d=1$, i.e. an assumed {\em a priori} rescaling of space and time that normalises the interaction range and diffusion coefficient to 1. Note that we set $L>2$, such that the boundaries cannot be simultaneously touched. We equip $[0,L]$ with periodic boundary conditions 
\[ u(0,t) = u(L, t), \qquad \partial_x u(0,t) = \partial_x u(L,t), \]
with the integral wrapped around in a natural way. The interaction strength parameter $\mu$ is taken as the bifurcation parameter.  

We define the Fourier-sine coefficients of the sensing function $\omega$ as 
\[ M_n(\omega) = \int_0^1 \sin\left(\frac{2\pi n r}{L}\right)\omega(|r|) dr. \]
As the monograph\cite{hillen2021} shows in detail, a number of properties can be identified for the following turning operator  
\[ K[u](x) = \int_{-1}^1 f(u(x+r,t)) \frac{r}{|r|} \omega(|r|) dr\,.\]
Specifically, $K$ is found to be skew adjoint, $K[1]=0$ and, for the specific case $f(u)=u$, maps sine and cosine functions as follows: 
\begin{eqnarray*}
    K\left[\sin\left(\frac{2\pi n x}{L}\right) \right] &=& 2 M_n(\omega) \cos\left(\frac{2\pi n x}{L}\right), \\ 
      K\left[\cos\left(\frac{2\pi n x}{L}\right) \right] &=& - 2 M_n(\omega) \sin\left(\frac{2\pi n x}{L}\right).      
\end{eqnarray*}
Moreover, if $u(x)$ is a steady state of (\ref{bif1}), then $u'(x) =0 $ if and only if $K[u]=0$,  $u''(x)\leq 0$ implies $K'[u]\leq 0$, and $K'[u]\geq 0$ implies $u''(x) \geq 0$. 
In this context we can view $K[u]$ as a non-local derivative and $K'[u]$ as a non-local curvature of the solution. 

The symmetries of $K[u]$ are also shown  \cite{hillen2021} to possess crucial properties. $K$ has $O(2)$ symmetry and, as a consequence, bifurcation branches arise at discrete points through the following theorem.
\begin{theorem}[see \cite{hillen2021}]\label{t:bif1} Consider a constant steady state   $\bar u$  of (\ref{bif1}) with $f'(\bar u)\neq 0$. For each $n=1,2,3,\dots$ with $M_n(\omega)>0$ there exists a bifurcation value and eigenfunction as 
\[ \mu_n = \frac{n \pi}{L\bar u f'(\bar u) M_n(\omega)}, \qquad e_n(x) =\cos\left(\frac{2\pi n x}{L}\right).\]
\end{theorem}
For a linear interaction function $f(u)=u$ it is also possible to identify the type of bifurcation via higher order expansions around the bifurcation value $\mu_n$. Specifically.
\begin{theorem}[see \cite{hillen2021}]
    If $f(u)=u$, then the type of bifurcation at $\mu_n$ is given by the sign of 
    \[ \beta_n = \frac{M_{2n}(\omega)-M_n(\omega)}{M_{2n}(\omega) - 2 M_n (\omega)}.\]
    If $\beta_n>0$ then the bifurcation at $\mu_n$ is supercritical and for $\beta_n<0$ it is subcritical. 
\end{theorem}
Notably, the type of bifurcation turns out to be entirely determined by the Fourier sine modes of the sensing function $\omega(r)$. \\

\noindent {\bf Example.}
As an example, consider $f(u)=u$ and a top-hat kernel 
\[ \omega(r) = \frac{1}{2}\chi_{[-1,1]}(r).\]
Then the Fourier sine coefficients of $\omega$ are 
\[ M_n(\omega) = \frac{L}{2\pi n}\sin^2\left(\frac{n\pi}{L}\right)\]
and the bifurcation values are 
\[ \mu_n = \frac{2\pi^2 n^2 }{L^2 \bar u \sin^2\left(\frac{n\pi}{L}\right)}. \]
If $L$ is a multiple of $\pi$, certain bifurcation values do not exist. Otherwise, all $\mu_n$ are well defined. The type of bifurcation is given by the sign of 
\[ \beta_n =2\left(1-\cot^2\left(\frac{n\pi}{L}\right)\right),\]
which, indeed, can be positive or negative. 

As in the previous subsection, a close relationship can be observed between model (\ref{bif1}) and those formulated from an energy based approach. Given the sensing function $\omega(r)$, we define a potential 
\begin{equation}\label{potential}
W(r) := V(r)\chi_{[-1,1]}(r), \qquad \mbox {with} \quad V'(r) = \omega(r) 
\end{equation}
Then for smooth solutions model (\ref{bif1}) is equivalent with 
\begin{equation}\label{bif2}
\partial_t u = \partial_{xx} u + \mu \partial_x [u \partial_x(W\ast f(u))]. \end{equation}
As such, the bifurcation result of Theorem \ref{t:bif1} can be straightforwardly extended to this case:
\begin{corollary} Consider (\ref{bif2}) where the potential is given by (\ref{potential}). Then,   for each $n=1,2,3,\dots$ with $M_n(\omega)>0$, there exists a bifurcation value and eigenfunction given by 
\[ \mu_n = \frac{n \pi}{L\bar u f'(\bar u)  M_n(\omega)}, \qquad e_n(x) =\cos\left(\frac{2\pi n x}{L}\right).\]
\end{corollary}

As a point of remark, in \cite{Carrillo2020} the bifurcations of (\ref{bif2}) were considered only for the linear case $f(u)=u$. For that case, the bifurcation value at equilibrium $\bar u = \frac{1}{L}$ was expressed as
 \[ \mu^*_n = - \frac{(2L)^{1/2}}{\tilde W(n)}, \]
where 
\[ \tilde W(n) = \sqrt{\frac{2}{L}} \int_0^L W(x) \cos\left(\frac{2\pi k x}{L}\right) dx\]
denotes the Fourier cosine coefficient of the potential $W$. We can directly compute that \[ \tilde W(n) = -\frac{\sqrt{2L}}{\pi n } M_n(\omega),\]
which implies $\mu_n = \mu^*_n$: a satisfying confirmation of our results. Note that in \cite{Carrillo2020} bifurcation analysis was also extended to arbitrary space dimensions, exceeding what has currently been performed for formulations of type (\ref{aggregation1}). 

\subsubsection{Weakly nonlinear analysis and conservation laws}

We observed above that bifurcations emerge at well-defined strictly positive wavenumbers, which is a rather typical behavior for   many reaction-diffusion systems  \cite{murray2003}. In most cases, weakly nonlinear analysis (WNLA) can  be used to reveal a Stuart-Landau equation governing the amplitude of the patterns close to the bifurcation point. However, when the PDE possesses a conservation law, i.e. $\frac{\rm d}{{\rm d}t}\int_\Omega u{\rm d}x=0$, the situation is rather more complicated.  In particular, the wavenumber that is destabilised first can be arbitrarily close to the origin, often meaning that the Stuart-Landau formalism is insufficient for capturing the dynamics of the amplitude of patterns   \cite{CH93, MC00}.  Such a situation is  pertinent here, as Equations (\ref{aggregation}) and (\ref{multiaggregation}) can all possess conservation laws under certain boundary conditions (e.g. periodic).  

To explain this in more detail, it is valuable to look at a specific example.  To this end, we consider a recently-studied symmetric 2-species version of Equation (\ref{multiaggregation})   given by\cite{giunta2023weakly}
	\begin{equation}\label{eq:ndsystem}
		\begin{aligned}
	        &\partial_t u_1= \partial_{xx} u_1+ \gamma\partial_{x} \left(u_1  \partial_{x}  (K \ast u_2) \right), \\
			&\partial_t u_2= \partial_{xx} u_2+ \gamma\partial_{x} \left(u_2 \partial_{x}  (K \ast u_1) \right),
		\end{aligned}
	\end{equation}
defined on $x \in \left[-\frac{L}{2},\frac{L}{2}\right]$ for $L>2$, with Supp$(K)=[-1,1]$ and periodic boundary conditions. Let $\bar{\bf u}=(\bar{u}_1,\bar{u}_2)$ be the constant steady state.  In the case $\gamma>0$, we can think of this as modelling two mutually-avoiding populations with identical advective and diffusive properties, for example territorial groups of animals.  For $\gamma<0$, this models mutually-attractive populations, for example symbiotic animal species, or cell-types that have mutual adhesive tendencies.

As is standard in WNLA, the authors\cite{giunta2023weakly} first decompose space and time into short and long scales.  Specifically, they define $X=\epsilon x$ and $T=\epsilon^2 t$.  Then  they look for solutions of the form\cite{MC00}
    \begin{equation}\label{eq:ansatz}
        \mathbf{u}(x,t)=\mathbf{\bar{u}} + {A}(X,T) e^{i q_c x} + {A^*}(X,T) e^{-i q_c x} + {B}(X,T),
    \end{equation}
where $q_c$ is the first wavenumber to be destabalised as $\gamma$ passes through the bifurcation threshold.  In situations where there is no conservation law, and the zero mode is stable close to the bifurcation point, there is no need to include the term ${B}(X,T)$.  However, the conservation law means that the zero mode always has an eigenvalue of zero so can be unstable to spatial perturbations on the slow-time, long-space scale (i.e. in $(X,T)$ coordinates). It should be noted that the amplitudes $ A$ and $B$ depend on the macroscopic time and space scales, while the mode $e^{i q_c x}$ depends on the microscale.
In particular, the authors showed \cite{giunta2023weakly} that if $\bar{u}_1 \neq \bar{u}_2$, $A$ is governed by the Stuart-Landau equation
\begin{equation}
\label{eq:Amplitude1}
A_T= q_c^2 A - \Lambda \lvert A \rvert^2 A,\\
\end{equation} 
and $B=0$, whenever $\gamma$ is in the linearly unstable regime.  However, if $\bar{u}_1 = \bar{u}_2$ then there is a different system of amplitude equations
\begin{align}
\label{eq:Amplitude2}
	&A_T= q_c^2 A - \Lambda \lvert A \rvert^2 A + \frac{q_c^2}{\bar{u}_1} A B,\\
	&B_T= \eta B_{XX}-\frac{1}{\bar{u}_1}( \lvert A \rvert^2)_{XX},
\end{align}  
where $\eta$ is a function of $\gamma_c$, $\bar{u}_1$, and $\hat{K}(0)$, where $\hat{K}(q)$ is the Fourier-cosine coefficient of $K(x)$
    \begin{equation}\label{eq:hatK}
        \begin{aligned}
	     \hat{K}(q)  = \int_{-1}^{1} K({x}) \cos( q x)\text{d} {x}.
        \end{aligned}
    \end{equation}
In Equations (\ref{eq:Amplitude1}) and (\ref{eq:Amplitude2}), $\Lambda$ controls the criticality of the bifurcation in $A$, and is a function of $\hat{K}(q_c)$, $\hat{K}(2 q_c)$, $\bar{u}_1$, $\bar{u}_2$, and $\gamma_c$ (see \cite{giunta2023weakly} for precise functional forms of $\Lambda$ and $\eta$).
In the $\bar{u}_1 = \bar{u}_2$ case, due to the contribution of the function $B(X,T)$, branches that bifurcate supercritically in $A(X,T)$ can be unstable.  Indeed, the following proposition holds.
\begin{proposition}\label{pr:stability}
     Suppose $\bar{u}_1=\bar{u}_2$. If $\gamma$ is in the linearly unstable regime and $ \Lambda>0$ then small amplitude patterns to System \eqref{eq:ndsystem} exist. These solutions are unstable if 
     \begin{equation}\label{eq:Gamma}
         	{\Lambda }<\frac{\bar{u}_1^2}{q_c^2\eta}.
     \end{equation}
\end{proposition}
This means that, in the case $0<\Lambda<\frac{\bar{u}_1^2}{q_c^2\eta}$, we have a supercritical bifurcation, but unlike the Stuart-Landau situation, stable patterns do not grow continuously as the bifurcation point is crossed.  Rather, numerical solutions show a discontinuous jump to a higher amplitude than the supercritical branch predicts\cite{giunta2023weakly}. This case study shows the importance of accounting for the zero mode in bifurcation analysis of nonlocal advection-diffusion equations.  Whilst we have only shown this in a single example, it is reasonable to expect that unstable supercritical branches may be a phenomenon observed more generally. 

\section{Discussion and Challenges}

To conclude, we outline a number of outstanding issues regarding modelling with nonlocal advection, and provide a few potential ways forward that could be fruitful in the coming years.

\medskip\noindent
{\textbf {Existence results.}} A large existence theory has been developed which covers a relatively broad spectrum of models that lie in the forms (\ref{aggregation}-\ref{multiaggregation}). However, the generalised structure of these equations can lead to a vast spectrum of models and an all-encompassing theory is not yet available. For example, functions ${\bf k}$ (or $w$) can vary from positive to negative and systems with multiple species can admit a wide spectrum of cross interactions. 

\medskip\noindent
{\textbf {Steady states, stability and bifurcation structure.}} Dynamically, models (\ref{aggregation}-\ref{multiaggregation}) are capable of an extremely rich variety of patterning, including stationary aggregate patterns, oscillating structures, travelling wave dynamics. Classical Turing-type stability analyses of nonlocal models have generally focused on one spatial dimension; intriguingly, however, recent extensions\cite{jewell2023} to higher dimensions indicate a dimensionally-dependent self-organising capacity, with patterning possible in higher dimensions for a formulation incapable of self-organisation in one-dimension. Studies into long time behaviours have primarily relied on simulations, however this alone is far from satisfactory: transients can persist over long timescales and become confused with stationary solutions. As an example, referring to Figure \ref{fig:linearstability}(f-g), a coarsening sequence is observed in which aggregates collapse over time: is the long time outcome a single aggregate? Expanding analytical methods, such as energy functional approaches \cite{carrillo2021phase,giunta2022detecting}, would have high value in generating a more nuanced understanding into steady states and bifurcation structures. 

\medskip\noindent
{\textbf {Boundary effects.}} In a nonlocal model, individuals inside some domain $\Omega$ may conceivably sense information from outside $\Omega$. The act of writing the nonlocal term at or close to a boundary therefore requires thought, as its support may extend beyond the domain of definition of the model. One can sidestep this through wrapping the nonlocal term around the domain, via the imposition of periodic boundary conditions\cite{giunta2022local}. Another approach is to alter the definition of the nonlocal term, in a mathematically consistent way, as it approaches the boundary\cite{hillen2020nonlocal}.  More broadly, the potential range of boundary conditions is immense and requires consideration on an application-to-application basis. For example, for adhesive populations one could allow the external space to exert varying levels of `stickiness’, or be actively repelling, according to tissue structure; in the case of multiple populations, different populations may respond distinctly near the interface. Non standard boundary conditions can strongly influence patterning within classical models (e.g., for reaction diffusion systems see \cite{dillon1994,Pottsedge}), and it is natural to expect a similarly powerful impact of boundary conditions on the aggregation models considered here.  

\medskip\noindent
{\textbf {Local formulations.}} Widescale adoption of nonlocal models is hindered by the analytical and numerical challenge. While efficient numerical methods have been developed -- Fast Fourier Transforms for the integral calculation \cite{gerisch2010}, positivity preserving finite volume methods \cite{carrillo2015}, pseudospectral methods \cite{goddard2022} --  formulating local models with similar properties could assist both numerics and  analysis. As noted, second order local models can be derived from random walk models \cite{anguige2009,johnston2012} and formal analyses\cite{eckardt2020} have investigated convergence between local and nonlocal forms. However, the potential of illposed local forms remains an issue. Fourth order local equations provide a promising avenue, and in \cite{falco2023} the following two species local model for sorting was derived from an underlying nonlocal system:
\begin{eqnarray*}
\partial_t u & = & -\nabla \cdot \left[ u \nabla \left( \mu \Delta u + \beta \Delta v +\gamma u + \delta v \right) \right]\,, \\
\partial_t v & = & -\nabla \cdot \left[ v \nabla \left( \beta \Delta u + \Delta v + \delta u + v \right) \right]\,.
\end{eqnarray*}
The parameters in the above relate to those in the nonlocal interaction terms and the above model was shown to be capable of reproducing a similar range of sorting dynamics to those of nonlocal models. Overall, derivation and exploration of well behaved local models is of importance. 

\medskip\noindent
{\textbf {Structured populations.}} Population heterogeneity in nonlocal models is typically restricted to two state systems, i.e. two populations with distinct properties. Discretisation into distinct subpopulations is often an approximation within biological systems: for example, studies\cite{kvokackova2021} of invasive breast cancer cells indicate invading cells lie on a continuum of intermediate states from epithelial to mesenchymal; individual-to-individual variation of `animal personality'\cite{ward2004,pettit2015,jolles2020} plays an important role in collective animal movements. Instead of extending the number of subpopulations in (\ref{multiaggregation}), subtle variation could be treated through a structured population framework: extending to a density $u({\bf x},p,t)$ where $p$ represents the phenotype state, and choosing interaction terms to describe how different phenotypes influence the dynamics \cite{perthame2006}. 

\medskip\noindent
{\textbf {Applications to sociological systems.}} This review has concentrated on nonlocal PDEs motivated by biological systems, in particular the spatiotemporal structuring of animals and cells. Naturally, the models and methods have applications beyond those areas, in particular to sociological systems. Perhaps the most germane example here would be crowds and traffic: an area that has witnessed much modelling with techniques ranging from agent-based to continuous (e.g. see \cite{bellomo2011,gong2023}). Concepts of stigmergy also cross to social systems, for example gang territoriality where agent-based modelling\cite{barbaro2013} has shown that territories can emerge indirectly through graffiti rather than direct conflict. Nonlocal models directly related to equation (\ref{aggregation1}) have been derived from agent-based models in the context of opinion dynamics (e.g. see \cite{balenzuela2015,garnier2017,pinasco2017,goddard2022}), where movement through physical space becomes a movement across opinion space and aggregation corresponds to consensus. Undoubtedly, numerous problems may benefit from the frameworks considered here.

\medskip\noindent
{\textbf {Testing predictions.}} Mathematical modelling of biological pattern formation is often inspired by the attempt to understand patterns already observed in biological systems: as examples, here we have described nonlocal models formulated to reproduce the observed patterns from cell sorting or territory formation.  However, one can also use models to predict patterns that could be observed. For example, the multi-species Equations (\ref{multiaggregation}) display rich pattern formation properties that ought to be observable in natural systems, if the models contain a sufficiently accurate representation of the underlying interactions. Patterns emerging from the model that have not yet been identified in the real world can be viewed as  predictions: do these patterns actually emerge in distributions if movement data is collected and/or analysed appropriately? If so, this would lead to new knowledge on the variety of patterns that can form spontaneously in populations of moving organisms. If not, this would inform us of missing features in our models, and deepen our understanding of the drivers of organism space use. 

\medskip\noindent
{\textbf {Connecting to data.}} Testing predictions demands techniques for connecting models and data. Beyond those reported here, as a further example, machine learning algorithms (e.g. see \cite{lu2019}) allow  trajectory data to be translated into interaction kernels for ABMs, which can then be scaled to PDEs. However, deciding the most appropriate for the data and question at hand is far from straightforward. To give an example from animal ecology, there are broadly two classes of techniques for fitting PDE models to data that are currently applied.  The first starts by building a PDE model based on qualitative aspects of behaviour that have been observed. Then the emergent patterns from numerical solutions of the PDE model are fitted to location data, to uncover the underlying behavioural processes in a more quantitative way.  This is exemplified in studies of mechanistic home range analysis\cite{moorcroftetal2006}.  The second approach follows that described in Section \ref{sec:ssa}, where a movement kernel (a.k.a. position jump process) like in Equation (\ref{eq:ssf}) is fitted to a time series of location data.  Then the PDE model is derived from this movement kernel \cite{potts2023scale}. The comparison between emergent pattern in the model and in the data then serves as a kind of `goodness-of-fit' test for the model, which can serve to uncover missing covariates of animal movement decisions \cite{potts2022assessing}.  Whilst this contrast in techniques has been known in the literature for some time\cite{potts2014animal}, these two approaches could do with some unification to achieve the maximum scientific benefit from analysing a given dataset.

\medskip\noindent
{\textbf {Collective cell movement.}} The analysis of collective cell movement forms a highly active area of research, from embryonic development to cancer invasion processes (e.g. see \cite{haeger2015,mayor2016}), and a large number of modelling approaches have been developed (e.g. \cite{Buttenschoen2020,alert2020}). Often, migrating cells extend long thin protrusions into their environment, possibly conferring an element of nonlocal sensing: for example, the formation of numerous lengthy filopodia appears to play an important role during effective migration of neural crest cells \cite{mclennan2020}, while long thin `tumour microtubes' play an apparently crucial role by facilitating invasion and growth of certain brain tumours (e.g. \cite{osswald2015,Microtubes}). Mathematical analysis of models capable of incorporating potential nonlocal impacts, as discussed here, promise new biological insight.

\medskip\noindent
The growth of mathematical biology in recent decades has been spectacular, crossing scales and disciplines. However, the trade off is  fragmentation: mathematical ecology, mathematical oncology etc. form their own fields, collaborative networks have become specialised, and keeping pace with developments in other fields becomes a challenge. Despite this,
the common language of mathematics remains. A key aim of this review has been to demonstrate this, showing the connection between nonlocal models used in ecological and cellular systems and suggesting the two fields can mutually benefit from their ongoing development.

\bigskip\noindent
{\bf Acknowledgements}: KJP is a member of INdAM-GNFM and acknowledges `Miur-Dipar\-timento di Eccellenza' funding to the Dipartimento di Scienze, Progetto e Politiche del Territorio (DIST). JRP acknowledges support of Engineering and Physical Sciences Research Council (EPSRC) grant EP/V002988/1. TH is supported through a discovery grant of the Natural Science and Engineering Research Council of Canada (NSERC), RGPIN-2017-04158.

\begin{appendix}
    \section{Correspondence between models}\label{a:integral} 
We demonstrate the calculations that show the translation between (\ref{ArmstrongNd}) and (\ref{energymin}). Specifically, we assume there exists a potential $W(|\br|)$ such that  
\begin{equation}\label{rightass} \nabla_\br W(|\br|) = \frac{\br}{|\br|} \omega(|\br|).
\end{equation}
Substituting (\ref{rightass}) into (\ref{ArmstrongNd}) and noting 
\[ \by = \bx+\br, \; \br=\by-\bx,\;  d\by = d\br,\;  \nabla_\by=\nabla_\br\,,\] 
\begin{eqnarray*}
u_t &=& d \Delta u -\mu \nabla\left(u\int_{B_R(x)} f(u(\bx+\br)) \nabla_\br W(|\br|) d\br\right) \\
 &=& d\Delta u -\mu \nabla \left(u\int_{B_R(0)} f(u(\by)) \nabla_\by W(|\by-\bx|) d\by \right)\\
 &=& d \Delta u +\mu\nabla\left( u \int_{B_R(0)} f(u(\by)) \nabla_\bx W(|\by-\bx|) d\by \right)\\
 &=& d\Delta u + \mu \nabla (u (\nabla_\bx W)\ast f(u) )\\[2ex]
 &=& d\Delta u + \mu \nabla (u \nabla(W\ast f(u))). 
 \end{eqnarray*}
 The above shows that energy minimisation corresponds to attractive interactions between individuals. Note that where subscripts are not included $\nabla \equiv \nabla _\bx$ and $\Delta \equiv \Delta_\bx$.

\end{appendix}
\printbibliography

\end{document}